\documentclass[sigconf]{acmart}

\AtBeginDocument{%
  }

\setcopyright{acmlicensed}
\copyrightyear{2018}
\acmYear{2018}
\acmDOI{XXXXXXX.XXXXXXX}
\acmConference[Conference acronym 'XX]{Make sure to enter the correct
  conference title from your rights confirmation email}{June 03--05,
  2018}{Woodstock, NY}

\acmISBN{978-1-4503-XXXX-X/2018/06}

\usepackage{url}
\usepackage{float}
\usepackage{amsmath}
\usepackage{algorithm}
\usepackage{algpseudocode}
\usepackage{tcolorbox}
\usepackage{enumitem}
\usepackage{multirow}
\usepackage{pifont}
\usepackage{xspace}
\usepackage{booktabs}
\usepackage{adjustbox}
\usepackage{ulem}
\usepackage{subcaption}

\DeclareMathOperator*{\myargmin}{arg\,min}
\DeclareMathOperator*{\myargmax}{arg\,max}

\newcommand{\attack}{VeFIT\xspace}
\newcommand{\method}{VeFIA\xspace}

\newcommand{\data}{$\mathcal{P}_d$\xspace} 
\newcommand{\task}{$\mathcal{P}_t$\xspace} 
\newcommand{\coord}{$\mathcal{C}$\xspace} 
\newcommand{\myparagraph}[1]{\noindent \textbf{#1}\xspace}
\newcommand{\preprocess}{\textbf{Stage1}\xspace} 
\newcommand{\train}{\textbf{Stage2}\xspace}
\newcommand{\infer}{\textbf{Stage3}\xspace}

\begin{document}

\title{On the Inference (In-)Security of Vertical Federated Learning: \\Efficient Auditing against Inference Tampering Attack}

\author{Chung-ju Huang}
\affiliation{%
  \institution{Key Laboratory of High-Confidence \\Software Technologies (MOE)\\ School of Computer Science \\ Peking University\\}
  \city{}
  \state{Beijing}
  \country{China}
}
\email{chongruhuang.pku@gmail.com}

\author{Ziqi Zhang}
\affiliation{%
  \institution{Department of Computer Science, University of Illinois Urbana-Champaign}
  \city{Champaign}
  \state{Illinois}
  \country{USA}}
\email{ziqi\_zhang@pku.edu.cn}

\author{Yinggui Wang}
\affiliation{%
  \institution{Ant Group}
  \city{Hangzhou}
  \state{Zhejiang}
  \country{China}
}
\email{wyinggui@gmail.com}

\author{Binghui Wang}
\affiliation{%
  \institution{Department of Computer Science, Illinois Institute of Technology}
  \city{Chicago}
  \state{Illinois}
  \country{USA}
}
\email{bwang70@iit.edu}

\author{Tao Wei}
\affiliation{%
 \institution{Ant Group}
 \city{Hangzhou}
 \state{Zhejiang}
 \country{China}
}
\email{lenx.wei@antgroup.com}

\author{Leye Wang}
\affiliation{%
 \institution{Key Laboratory of High-Confidence \\Software Technologies (MOE)\\ School of Computer Science \\ Peking University\\}
 \city{}
 \state{Beijing}
 \country{China}}
\email{leyewang@pku.edu.cn}

\renewcommand{\shortauthors}{Huang et al.}

%
\begin{abstract}
Vertical Federated Learning (VFL) is an emerging distributed learning paradigm for cross-silo collaboration without accessing participants' data.
However, existing VFL work lacks a mechanism to audit the inference correctness of the data party. The malicious data party can modify the local data and model to mislead the joint inference results.
To exploit this vulnerability, we design a novel \underline{Ve}rtical \underline{F}ederated \underline{I}nference \underline{T}ampering (VeFIT) attack, allowing the data party to covertly tamper with the local inference and mislead results on the task party's final prediction.
VeFIT can decrease the task party's inference accuracy by an average of 34.49\%.
Existing defense mechanisms can not effectively detect this attack, and the detection performance is near random guessing.
To mitigate the attack, we further design a \underline{Ve}rtical \underline{F}ederated \underline{I}nference \underline{A}uditing (VeFIA) framework. VeFIA helps the task party to audit whether the data party's inferences are executed as expected during large-scale online inference. VeFIA does not leak the data party's privacy nor introduce additional latency.
The core design is that the task party can use the inference results from a framework with Trusted Execution Environments (TEE) and the coordinator to validate the correctness of the data party's computation results.
VeFIA guarantees that, as long as the proportion of inferences attacked by VeFIT exceeds 5.4\%, the task party can detect the malicious behavior of the data party with a probability of 99.99\%, without any additional online overhead.
VeFIA's random sampling validation of VeFIA achieves 100\% positive predictive value, negative predictive value, and true positive rate in detecting VeFIT.
We further validate VeFIA's effectiveness in terms of privacy protection and scalability on real-world datasets.
To the best of our knowledge, this is the first paper discussing the inference auditing problem towards VFL.
We provide the artifact at \url{https://anonymous.4open.science/r/VeFIA}.
\end{abstract}

\ccsdesc[500]{Computing methodologies~Distributed artificial intelligence}
\ccsdesc[500]{Security and privacy}

\keywords{Vertical federated learning, Inference auditing, Trusted execution environments}

\received{20 February 2007}
\received[revised]{12 March 2009}
\received[accepted]{5 June 2009}

\maketitle

\section{Introduction}

\myparagraph{Vertical Federated Learning (VFL).}
Federated Learning (FL) enables multiple participants to jointly train a machine learning model without sharing private data~\cite{YangLCT19,abs-2304-01829,McMahanMRHA17}, naturally fitting cross-enterprise data collaboration where multiple companies jointly extract value from shared workflows while keeping proprietary data local.
Vertical FL (VFL) is a type of FL in which two participants hold \textit{different features} for the \textit{same users}.
The participant holding the labels is the \textit{task party} \task and the other is the \textit{data party} \data~\cite{YangLCT19}. 
VFL also includes a third-party coordinator~\cite{YangLCT19,abs-2304-01829} to coordinate the communication process. 
The coordinator (abbreviated as \coord or COO) in VFL is typically an FL service platform, e.g., TensorOpera~\cite{TensorOpera}, which can provide services such as cloud inference, resource matching, and scheduling.
VFL is widely used in scenarios such as e-commerce, healthcare, and
finance~\cite{LiuKZPHYOZY24,abs-2405-02364,GuLKFY23,HuangW023}.
For example, in financial risk control (e.g., real-time credit-card fraud detection), the task party \task can be an issuing bank or a FinTech lender that holds labels together with its own transaction features~\cite{abs-2009-06218,ChengLCY20}.
The data party \data can be a payment service provider that owns complementary behavioral and device-side features for the same users.
\task collaborates with \data via VFL to jointly train a fraud scoring model without exchanging raw features.
The business model of VFL is that \task asks \data to participate in training and pays \data during online inference based on the amount of data queried (i.e., pay-as-you-go).
This paper focuses on the \textbf{inference phase} in VFL, where the system serves high-throughput and latency-sensitive inference requests~\cite{SoiferLLZLHZOMB19, GujaratiKAHKVM20}.
This setting is particularly challenging in real-world VFL deployments under strict service-level latency constraints.


\myparagraph{Lack of Inference Auditing.}
In cross-enterprise collaboration, the core difficulty is not only privacy but also trust~\cite{responsible-ai}: parties must be confident that collaborators follow the agreed protocol and do not tamper with data or models.
In the VFL inference phase, \data receives the inference request from \task and uses the target data and the trained model for inference.
Although VFL keeps raw data local to mitigate privacy concerns, \textit{existing VFL lacks inference auditing for \data}, providing no end-to-end trust guarantee.
Thus, \task cannot ascertain the legitimacy of the data or model used by \data, nor the integrity of the \data's inference process.
During the inference phase, malicious \data can modify its data and model, and generate an incorrect representation.
\coord aggregates the representations and forwards the results to \task.
\task uses unreliable aggregation results to make final predictions, leading to degraded user experience and potential economic losses.
\textit{To the best of our knowledge, this is the first work to discuss the problem of inference auditing and its security consequences in VFL.}

\myparagraph{Our Attack.}
To exploit this vulnerability, we propose a novel \underline{Ve}rtical \underline{F}ederated \underline{I}nference \underline{T}ampering (\attack) attack, which is complementary to prior works.
The attack enables \data to mislead \task's joint inference by tampering with local models and data.
The attack consists of three stages: offline surrogate model training, adversarial noise generation, and online attack.
\data (attacker) first trains a surrogate model to mimic the behavior of \task's model.
Then, \data generates two types of adversarial noises: model and data noise.
The model noise is added to the local model, and the data noise is added to the local data.
Finally, when \data wants to conduct the attack, it can use the adversarial noises to generate harmful representations.
Such representation lies in a similar space of benign representations but can mislead the model to make wrong predictions.
Note that the noises are input-agnostic and apply to all samples to generate the adversarial representations.
Experimental results demonstrate that \attack can degrade the accuracy of \task's joint inference by an average of 34.49\%.
Existing defenses~\cite{LaiW0LZ23,ChenZLGW24,ChoHYLBP24}, which rely on statistical representation differences, fail to detect \attack, achieving only 52.53\% Positive Predictive Value (PPV), 53.46\% Negative Predictive Value (NPV), and 29.28\% True Positive Rate (TPR) on average--comparable to random guessing.


\myparagraph{Defense Insight.}
To address the lack of inference auditing problem and defend against the \attack, we propose a \underline{Ve}rtical \underline{F}ederated \underline{I}nference \underline{A}udit framework, \textbf{\method}, based on Trusted Execution Environments (TEE).
TEE ensures that the computation is isolated and protected from untrusted systems.
In VFL, participants are typically enterprises and institutions with sufficient computing resources, making the deployment of TEE feasible and practical.
The core of inference auditing is to generate verifiable evidence to assess whether \data's inference adheres to expected standards.
Our insight is that \data leverages the collaborative inference between \data's TEE and outsources partial computation to the coordinator \coord.
The TEE-COO trusted collaborative inference runs in parallel with \data's untrusted inference, providing trusted results that serve as evidence to validate the correctness of \data's inference. 
\task can utilize this evidence to determine whether \data's inferences have been modified.
The goal of \method is to \textit{enable \task to audit whether \data's inferences are performed as expected, without compromising data privacy of \data or introducing extra latency}.

\myparagraph{Technical Challenges.}
However, there are three technical challenges in designing an efficient
auditing framework based on TEE.
%
%
\textbf{C1: Privacy Leak Measurement.}
The first challenge is how to mitigate the privacy leakage of inference outsourcing.
Previous research has shown that outsourcing part of the local model to the cloud for inference can leak the owner's input privacy~\cite{HeZL19,YinZZLYCH23,PasquiniAB21}.
Given that the primary objective of VFL is to preserve participant privacy, it is crucial to ensure that the input privacy of the \data remains protected throughout the TEE-COO collaborative inference.
\textbf{C2: Runtime Subject Authenticity.}
The second challenge is how to ensure the authenticity of the data and model used in TEE-COO collaborative inference.
Existing mechanisms that leverage TEE to validate outsourced computations on untrusted accelerators (matrix multiplication~\cite{Zhang0ZZZ0W24} and model mutation~\cite{sun2023shadownet}) are inadequate in this context, as they presume the data and model to be authentic and do not verify the correctness.
%
\textbf{C3: Validation Efficiency.}
The third challenge is how to efficiently audit the untrusted inference during the inference stage with a large volume of queries.
The speed of TEE-COO trusted collaborative inference is significantly slower than that of \data’s untrusted accelerator~\cite{LiZGC0ZG21,LeeLPLLLXXZS19}. Auditing all samples will introduce unacceptable overhead.


\myparagraph{Our Solution.}
To address these challenges, our \method consists of three components: privacy-aware training, runtime authenticity validation, and efficiency-aware computation schedule.
To solve \textbf{C1}, we use mutual information to measure the privacy knowledge of the original data of \data in the outsourced inputs.
During training, we use privacy-aware training based on information theory~\cite{AlemiFD017,belghazi18a,ChengHDLGC20} to reduce privacy knowledge and maintain the overall training performance of VFL.
Our theoretical analysis (Thm.~\ref{theorem:bound}) shows that lowering the mutual information between the outsourced inputs and the private features provably increases the minimum privacy reconstruction error of any inversion attacker.
%
To solve \textbf{C2}, the second component validates the data and models loaded in TEE based on hash consistency.
Data validation ensures that the runtime inference data hash matches its stored version, while model validation ensures that the runtime model hash matches the model version that has been performance-tested prior to inference.
To solve \textbf{C3}, the third component confidentially and randomly selects a subset of inference samples for validation via the TEE-COO pipeline, thereby maintaining efficiency during large-scale inference.
The selective auditing mechanism enables \task to detect malicious behavior with high probability (e.g., 99.99\%) without incurring additional latency once the fraction of malicious inferences surpasses a minimal threshold~\cite{zhao2021veriml,huang2022zkmlaas,zhou2021zero}.
To further improve performance, we integrate a pipeline acceleration strategy and enhance the overall efficiency of TEE-COO inference.

\myparagraph{Evaluation.}
We extensively evaluate the effectiveness and efficiency of \method.
The evaluation covers five datasets, three models, nine metrics, and three defense baselines.
Experimental results demonstrate that when the proportion of malicious inferences exceeds 5.4\%, \method can detect \attack from \data with a probability of 99.99\%.
In random sampling-based validation, \method achieves 100\% PPV, TRP, and NPV.
Moreover, privacy-aware training reduces the risk of input data leakage from the \data by an average of 72.9\% and the accuracy drops by only 0.51\%.
\method introduces only a negligible overhead of 0.56\%.
Besides, the pipeline acceleration for TEE-COO collaborative inference can reach an average of 4.38$\times$.
Finally, even when the number of \data increases, the fluctuation in detection time is slight (averagely 5\%).

\myparagraph{Contributions.}
We summarize the contributions below:
\begin{itemize}[leftmargin=*]
    \item To the best of our knowledge, this is the first paper to study the lack of inference auditing problem in VFL.
    \item We propose \attack, a novel inference tampering attack against \task in the VFL inference stage.
    \item We design \method, an inference auditing framework, to efficiently validate the trustworthiness of the \data's inference.
    \item Extensive experimental results show \method can effectively detect the inference-time attack with no additional system overhead.
    
\end{itemize}

\section{Background}

In this section, we provide the background for this paper, including an overview of the VFL pipeline and problem scenario (Sec.\ref{subsec:vfl}).
Next, we introduce the TEE-Coordinator (TEE-COO) framework (Sec.\ref{subsec:tee_coo}), which is widely adopted in FL deployments.
For clarity, we summarize the key symbols in Tbl.~\ref{tab:notation}.

\subsection{Vertical Federated Learning}
\label{subsec:vfl}
\begin{figure}[!t]
\centerline{\includegraphics[width=0.85\linewidth]{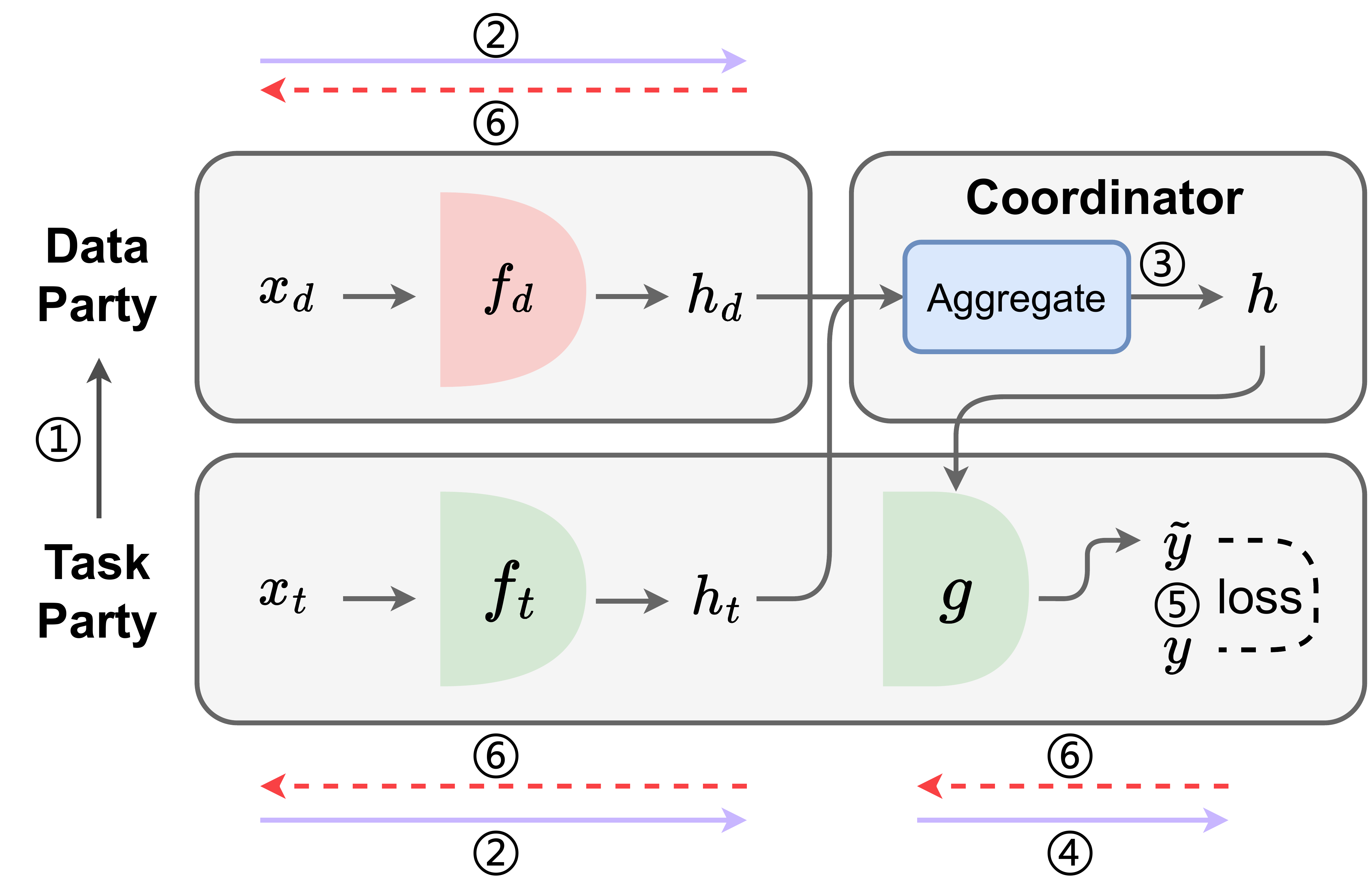}}
\caption{The pipeline of the typical VFL process. Purple solid and red dashed arrows represent forward and backward dataflow, respectively.
}
\Description{The training/inference pipeline of splitVFL.}
\label{fig:splitvfl}
\end{figure}

VFL is widely applied in cross-silo collaborative tasks in healthcare, finance, and e-commerce~\cite{NguyenPPDSLDH23,QiuZJDPZW23,ZhangG22}.
In a typical VFL scenario, there is one task party \task and one data party \data, and also a third-party coordinator \coord to assist the training and inference~\cite{PangYSW23,LuoWXO21,WangG0Z23,YangLCT19}.
The VFL task is initiated by \task.
For each sample $(x,y)$ with features $x$ and label $y$, \task and \data hold its respective features $x_t$ and $x_d$ of the sample such that $x=[x_t;x_d]$.
\task also holds the label $y$.
Due to privacy concerns, \data and \task keep their raw features locally, but share extracted high-level features with others.
To do so, \task and \data independently train their bottom model (i.e., a feature extractor), denoted as $f_t$ and $f_d$.
Then \task and \data send the outputs of the bottom models $h_t=f_t(x_t)$ and $h_d=f_d(x_d)$ to \coord.
\coord combines $h_t$ and $h_d$ to construct an aggregated representation $h$, and sends $h$ to \task, which uses $h$ to train a top model $g$ to predict a label that matches the ground truth label $y$.
Fig.~\ref{fig:splitvfl} overview the VFL framework~\footnote{Appx.~\ref{app:background} shows an illustrative comparison between VFL and Horizontal FL (HFL).}.

\begin{table}[htbp]
\caption{Key Symbols used in this paper.}
\begin{center}
\adjustbox{max width=0.95\linewidth}{
\begin{tabular}{ccc}
\toprule
\textbf{Type} & \textbf{Notation} & \textbf{Description} \\ \midrule
\multirow{4}{*}{Common} & \task,\data, \coord & Task party, data party, coordinator \\
& $f_t, g, x_t$ & \task's bottom model, top model, private data \\
& $y, \tilde y$ & ground-truth labels and predicted labels \\
& $f_d, x_d$ & \data's bottom model and private data \\
& $h_t,h_d$ & Local representations held by \task and \data \\ \midrule
\multirow{4}{*}{\attack} & $g_s$ & Surrogate top model \\
& $x_d^{aux},y_d^{aux}$ & Auxiliary samples for training attack models \\
& $\hat{x}_d,\hat{f}_d,\hat{h}_d$ & Attacked $x_d$, $f_d$ and $h_d$ \\
& $\delta_x,\delta_f$ & Perturbation noises for \attack \\ \midrule
\multirow{7}{*}{\method} &$f_d^{sm},f_d^{dm}$ & \data's shallow and deep bottom models \\
& $\sigma$ & Perturbations on $z_d$ \\
& $z_d,\hat{z}_d$ & Original and perturbed outputs of $f_d^{sm}$ \\
& $N, W$ & Number of inferences and sampling validation ratio \\
& $W_*$ & The optimal sampling validation ratio. \\
& $K$ & The proportion of malicious tampering inference \\
& $T_{un},T_{tr}$ & Running time of untrusted/trusted inference. \\
\bottomrule
\end{tabular}
}
\label{tab:notation}
\end{center}
\end{table}

\myparagraph{Pipeline.}
A typical VFL consists of three stages: preprocessing (\preprocess), joint training (\train), and joint inference (\infer). 

\noindent $\bullet$ \preprocess.
As a common practice in existing VFL deployments~\cite{PangYSW23,LiuKZPHYOZY24,LuoWXO21}, \task and \data employ the Private Set Intersection (PSI)~\cite{FreedmanNP04} to identify the aligned sample IDs without revealing data information.
Before running PSI, all parties agree on a canonical entity key (e.g., account/device ID) and a deterministic canonicalization function $\text{Canon}(\cdot)$ (including encoding and normalization rules).
The PSI output is a set of cryptographic tokens $H(\text{Canon}(\text{key}))$, which serve as the globally consistent aligned sample IDs.
\task then partitions these aligned IDs into three subsets: training, validation, and test.
Note that \task possesses the ground-truth labels for the training and validation sets, but not for the test set--these labels must be predicted using the trained VFL model.


\noindent $\bullet$ \train.
The training phase consists of six iterative steps (\ding{172} to \ding{177}), as illustrated in Fig.~\ref{fig:splitvfl}.
In the figure, purple solid arrows denote forward data flow, while red dashed arrows denote backward propagation.
At the beginning of each iteration (\ding{172}), the \task sends the IDs of the training samples to be used in that iteration to the \data.
Both \data and \task uses the IDs to load $x_d$ and $x_t$, and use the bottom models ($f_d$ and $f_t$) to compute the extracted representation $h_d = f_d(x_d)$ and $h_t=f_t(x_t)$ (\ding{173}).
These representations are then transmitted to the \coord, who concatenates them to form a joint representation $h=[h_d;h_t]$ (\ding{174}) and sends it to the \task.
\task uses its top model $g$ to compute the predicted label $\tilde{y}$ (\ding{175}), and evaluates the loss $l(y,\tilde{y})$ (\ding{176}).
Subsequently, \task computes the gradient of the loss and sends the gradient with respect to $h_d$ to the \data.
Both \data and \task then use the gradients to update their respective local models $f_t$, $f_d$, and $g$.
\task monitors the model's performance on the validation set to determine when to terminate training--e.g., upon reaching a predefined accuracy or maximum number of iterations.

\noindent $\bullet$ \infer.
VFL inference follows the ``Pay-as-You-Go'' practice~\cite{abs-2304-01829,PangYSW23}, where \task pays to \data based on the volume of data utilized.
The inference process includes all the forward steps (\ding{172} to \ding{175}) for test samples.
Specifically, \task sends the IDs of the test samples to \data and pays the fee proportional to the number of test samples.
Subsequently, the \task receives the aggregated representations from all parties via the \coord and performs the final prediction using $g$.



\myparagraph{Scenario.}
We consider the online joint inference phase (\infer) in VFL.
Take real-time credit-card fraud detection as an example: when a user initiates a transaction, the issuing bank \task needs to decide whether to approve or decline it within a strict latency budget.
\task leverages both its own financial features and the complementary risk signals held by \data through VFL.
During traffic spikes (e.g., promotional campaigns or shopping festivals), the system can receive a burst of transactions, making the online inference pipeline highly throughput-sensitive. It is important to practically audit \data's inference behaviors efficiently.

\subsection{VFL Security}

Existing attacks on VFL include \textit{privacy attacks}~\cite{Fu0JCWG0L022,WangLZZSRL23,YangMZKKR24,LuoWXO21,QiuZJFYW24} and \textit{training-time attacks}~\cite{LaiW0LZ23,ChenZLGW24,naseri2023badvfl,BaiCZ0WG23,ChoHYLBP24}.
In privacy attacks, the attacker adheres to the VFL protocol but attempts to infer private data information.
Training-time attacks (e.g., backdoor attacks) manipulate the training data in \train to decrease the overall test accuracy or implant hidden vulnerabilities into the model. 
Note that training-time attacks require changing features for a long time to attack the training process consistently.
This long-term attack leads to significant statistical changes to intermediate parameters (e.g., embeddings and gradients), and is more susceptible to be detected by anomaly detection mechanisms~\cite{ChoHYLBP24,ChenZLGW24,LaiW0LZ23}.
Our work presents a practical inference-time attack that complements prior works.

Several studies have explored detection mechanisms for training-time attacks on VFL~\cite{LaiW0LZ23,ChenZLGW24,ChoHYLBP24}
These defenses rely on using deep learning techniques (e.g., contrastive learning~\cite{abs-1807-03748}, masked auto-encoder~\cite{HeCXLDG22}) to perform anomaly detection on the statistical properties of intermediate representations (e.g., $h_d$ of \data).
Since training-time attacks require consistent malicious injections, and will cause significant statistical deviations in intermediate representations, making them more susceptible to detection.
However, our attack, \attack (introduced in Sec.~\ref{sec:attack}), faithfully performs the training phase and does not modify training data.
Thus our attack is more stealthy and will not be detected by existing defenses.

\textbf{Limitation of Prior Defense.}
Most of the defense mechanisms against training-time attacks on FL are designed for HFL~\cite{SandeepaSWL24,KabirSRM24,MozaffariSH23,FereidooniPRDS24,ChuGISL23,ShejwalkarH21,KraussD23}.
Since the training paradigms and architectures of VFL and HFL are significantly different, the defense mechanisms in HFL cannot be applied to VFL~\cite{BaiCZ0WG23,naseri2023badvfl}.

\subsection{TEE-COO Framework}
\label{subsec:tee_coo}
\myparagraph{Characteristic.}
The TEE-COO framework has been widely applied in FL~\cite{MoHKMPK21,Kato0Y23,ZhangLZLWW20}. 
We leverage the TEE-COO collaborative inference mechanism in VFL to assist \task in auditing the inferences of \data.
This mechanism is a widely used framework for securing inference in the multi-party ML system~\cite{abs-1912-03485,mo2020darknetz,Zhangteeslice2024}.
It can be viewed as an extension of edge-cloud collaborative inference, where the edge (e.g., \data) is equipped with a TEE, and the cloud serves as the coordinator \coord.
The service provider can run the validation programs in the TEE to ensure the trustworthiness of the inference results.
The characteristic of TEE is its ability to isolate a secure enclave from the edge environment, ensuring that computations within the enclave are protected from manipulation by the untrusted edge.
But it has limited computational resources and comparatively low inference (sometimes the
performance overhead is 6-7$\times$ slower than the native inference)~\cite{LiZGC0ZG21,LeeLPLLLXXZS19}.
The characteristic of \coord is that it has high computational resources but may
not be fully trusted by \data.
Thus it is suitable for running the computation-intensive but non-sensitive part of the inference program~\cite{abs-1912-03485,mo2020darknetz}. 
\data divides its bottom model $f_d$ into a shallow bottom model $f_d^{sm}$ and a deep bottom model $f_d^{dm}$, which are placed in its local TEE and the remote \coord, respectively, for collaborative inference (see Fig.~\ref{fig:model_partition} in Appx.~\ref{app:tee_coo}).
In FL, models are often treated as tradable assets~\cite{LiLLX23,FengHLNJXKYNM22}, meaning they can be sent to third-party servers for auditing and validating~\cite{WangCMMW24,ZhangY22,PengXCGYGT22}.

\myparagraph{Advantages.}
TEE-COO has several advantages over prior frameworks. 
Compared with the edge-cloud framework, TEE-COO can defend against the attack from edge owners or third-party attackers on the edge~\cite{sun2023shadownet,Zhangteeslice2024} by running validation programs in the TEE.
Compared with the TEE-only framework, TEE-COO can reduce the computation overhead of the TEE by offloading the computation-intensive part to \coord~\cite{abs-1912-03485,ShenQJWWCZWCLZC22,TramerB19}. 
Compared with \coord-only framework, TEE-COO can protect data privacy by running the inference program on the edge to comply with the privacy protection of VFL.




\section{Threat Model}
\label{sec:threat_model}

\myparagraph{Adversary Goal.}
We consider the data party (\data) to be the adversary that aims to reduce the task party's (\task) inference accuracy of \infer during the joint inference.
Our attack targets the inference phase (\infer), rather than the preprocessing (\preprocess) or training (\train) phases.
Specifically, \data is \textit{semi-honest}~\footnote{A semi-honest entity follows the protocol correctly but attempts to infer private information from the received data--a standard setting in FL~\cite{YangLCT19,GaoZ23,PangYSW23}.} during \preprocess and \train, but acts maliciously during \infer~\cite{PangYSW23,LuoWXO21}.
The goals of \data are three-fold:

\noindent
$\bullet$ \textbf{AG1: Effectiveness.} 
\data may attempt to tamper with their test samples or the bottom model to mislead the final prediction of \task's top model and reduce the \task's test accuracy.


%
\noindent
$\bullet$ \textbf{AG2: Stealthiness.}
The modification in the representation space before and after tampering should be as minor as possible.
If the modifications are substantial, the distance between the benign and adversarial representations could be large, making them detectable by similarity-based defenses~\cite{ChoHYLBP24,ChenZLGW24,LaiW0LZ23}.

\noindent
$\bullet$ \textbf{AG3: Efficiency.}
During \infer, stringent per-sample latency requirements necessitate that online generation of adversarial samples is highly efficient.
If the process is too slow, it can cause request congestion and degrade VFL inference throughput.

\myparagraph{Adversary Capability.}
In VFL, each participant only has access to its subset of features and cannot access other participants'.
Consequently, \data has full control over $x_d$ and $f_d$ and may modify them arbitrarily, but cannot access or alter \task's data, including $x_t$, $f_t$, and $g$.
During \train, \data honestly follows the VFL training protocol and therefore observes its local representation $h_d$ and gradient from \task.
Prior work has shown that such training traces are sufficient to recover a fraction of training labels via label inference attacks in VFL~\cite{Fu0JCWG0L022,yang2024learning}.
Following these work, we assume that \data can obtain a small auxiliary labeled set $\{x_d^{aux}, y_d^{aux}\}$, where $x_d^{aux}$ is sampled from \data's \emph{own} local feature vectors (i.e., the $P_d$-side features of some aligned training IDs) and $y_d^{aux}$ are labels inferred from the observed training traces. 
Based on this auxiliary set, \data locally trains a surrogate top model $g_s$ that approximates the prediction behavior of $g$ on \data's feature space.
Importantly, we do not require \data to actively query the \task's deployed model in \infer, nor to have access to the same training data as \task; it only relies on information that is already available to \data during \train.



\myparagraph{TEE Security.}
We assume TEE is available on \data's machine following prior work~\cite{Zhang0ZZZ0W24,Zhangteeslice2024}.
\task utilizes the TEE on \data's machine to defend against potential attacks.
We follow prior work to adopt three security features provided by TEE: remote attestation, encryption, and integrity~\cite{sun2023shadownet,Zhangteeslice2024}.
Remote attestation enables \task to verify and securely execute a trusted program within the TEE enclave.
Encryption ensures that the TEE's data is encrypted, and \data cannot see the data
in TEE.
Integrity guarantees that \data cannot modify the program nor the data in TEE.
Additionally, TEE can seal data and store the sealed data on disk.
The sealed data is encrypted, and \data cannot tamper with it because every time TEE loads the sealed data, it will check the integrity.
Besides, side-channel attacks~\cite{abs-1811-05378,GaoQZWMAXFN24,LiWW0TZ22} against TEE are beyond the scope of this paper, in line with assumptions made in prior literature~\cite{RiegerKMDS24,Zhangteeslice2024,mo2020darknetz,HouLLWWL22}.


\myparagraph{Coordinator.}
The coordinator \coord provides high-performance computing resources and is modeled as \textit{semi-honest}~\cite{YangLCT19,HeZL19,GaoZ23}. Specifically, while \coord honestly executes the required computational tasks during \train and \infer, it acts as a curious observer that may attempt to infer private information of \data or \task from the received intermediate representations.
Consistent with prior work~\cite{YangLCT19,XuB00JL21,HeZL19,GaoZ23}, \coord is considered as an independent third party that does not collude with \data or \task, as such collusion would severely undermine its business credibility and neutrality.
We discuss the implications of potential collusion and future mitigation strategies in Sec.~\ref{sec:dis}.

\myparagraph{Defender's Goal.}
\task is the defender and aims to detect malicious inference during \infer by validating the transmitted representations $h_d$.
When \task identifies unexpected inference results, \task can halt the joint inference and reclaim the inference costs.
During \infer, the high frequency of inference requests may prevent \task from validating every result.
To mitigate this, \task seeks to \textit{validate as many inference results as possible without introducing additional latency}, thereby limiting \data's ability to launch a large-scale attack.
As a result, any increase in the number of tampered inferences significantly raises \task's likelihood of detecting the attack.

\myparagraph{Defender's Capability.}
\task's capabilities align with prior VFL work~\cite{QiuZJFYW24,QiuZJDPZW23}.
It faithfully executes the VFL pipeline and controls $x_t$, $f_t$, and $g$, but is not permitted to directly access \data's $x_d$ or $f_d$ for auditing purposes.
\task can remotely attest the auditing program in \data's TEE.
\task does not have control over \coord and relies solely on the inference results received from \coord for auditing.

\section{Inference Tampering Attack}
\label{sec:attack}
\begin{figure}[!t]
\centerline{\includegraphics[width=0.9\linewidth]{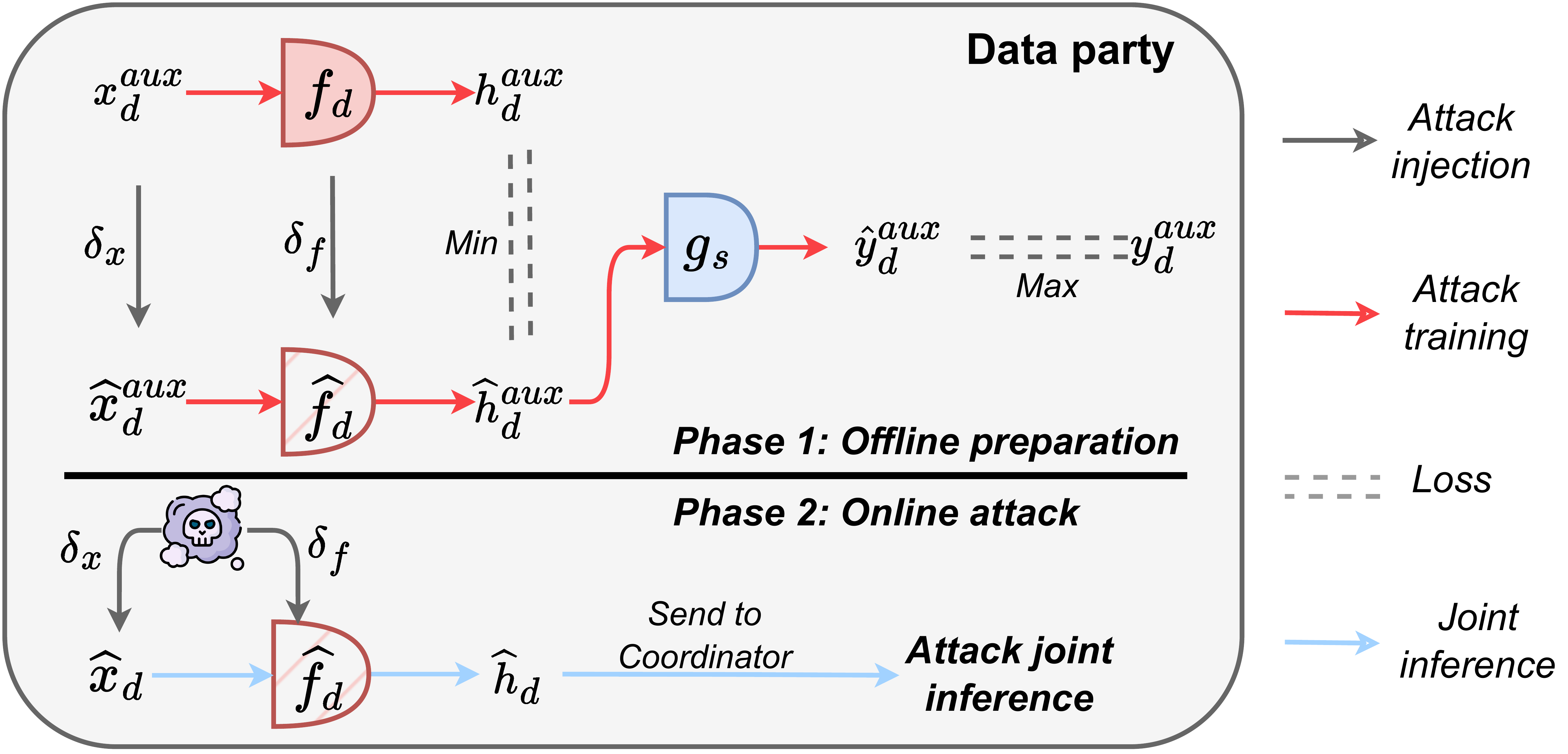}}
\caption{The pipeline of \attack.
In the offline preparation phase, \data trains the adversarial noises $\delta_x$ and $\delta_f$.
In the online attack phase, \data uses $\delta_x$ and $\delta_f$ to tamper with local inference and attack joint inference of \task.
}
\Description{}
\label{fig:attack}
\end{figure}

In this section, we introduce our inference tampering attack, \attack (\underline{Ve}rtical \underline{F}ederated \underline{I}nference \underline{T}ampering) launched in the inference stage \infer.
We first formulate the attack (Sec.~\ref{subsec:attack_formulation}) to consider the attack goals (\textbf{AG1}-\textbf{AG3}) and then present the attack design (Sec.~\ref{subsec:attack_design}).
We then illustrate the evaluation setup (Sec.~\ref{subsec:attack_eval_setup}) and evaluate the attack (Sec.~\ref{subsec:attack_evaluation}).
Fig.~\ref{fig:attack} shows the \attack pipeline.

\subsection{Attack Formulation}
\label{subsec:attack_formulation}


Recall the adversary \data completely controls its data $x_d\sim D_d$ and bottom model $f_d$.
\data can easily train two noises $\delta_x$ and $\delta_f$, which are added to $x_d$ and $f_d$, respectively. 
For efficiency purposes (\textbf{AG3}), \data does not require retraining the noises for new test samples.
Instead, $\delta_x$ may be applied across multiple samples in \infer.
Let $\hat{x}_d$ and $\hat{f}_d$ be the tampered input and model, respectively.
The adversarial representation $\hat{h}_d = \hat{f}_d(\hat{x}_d)$ is then sent to \coord for aggregation.
$Dist(\cdot)$ is a distance function that applies to the representation space, and $\delta_h$ is the feature distance constraint.
The attack goal is to mislead the prediction of \task's top model $g$ (\textbf{AG1}) while minimizing the modification in the representation space (\textbf{AG2}).
This goal can be formulated as follows:

\begin{equation}
\begin{aligned}
    \myargmax_{\delta_x,\delta_f}\sum_{x_d\sim D_d}&\mathbb{I}(g(\hat{h}_d,h_t) \neq g(h_d,h_t))\\
    \text{s.t.} \quad {Dist}(h_d, \hat{h}_d) < \delta_h &, \quad \hat{h}_d = \hat{f}_d(\hat{x}_d), \quad \hat{x}_d = x_d + \delta_x,
\end{aligned}
\end{equation}
where $\hat{f}_d$ is a perturbed model by adding $\delta_f$ to $f_d$, and $\mathbb{I}(\cdot)$ is the indicator function.



\subsection{Attack Design}
\label{subsec:attack_design}
The \attack's design consists of two phases: offline optimization and online attack.
In the offline phase, \data trains a surrogate model $g_s$ and adversarial noises $\delta_x$ and $\delta_f$.
In the online attack phase, \data constructs adversarial samples $\hat{x}_d$, performs the inference on $\hat{f}_d$, and sends $\hat{h}_d$ to \task.
Alg.~\ref{alg:attack} and Fig.~\ref{fig:attack} show the attack pipeline.



\begin{algorithm}[!t]
\small
    \caption{Inference Tampering Attack}
    \label{alg:attack}
    \begin{algorithmic}[1]
        \Procedure{OfflinePreparation}{$x_d^{aux},y_d^{aux},f_d$}
            \State Initialize $g_s$ 
            \While{not converged} \Comment{Surrogate Model Training}
                \State $y_d^{pred}\gets g_s(f_d(x_d^{aux}))$
                \State $g_s\gets \mathit{Backward}(\mathit{CrossEntropy}(y_d^{pred},y_d^{aux}))$
            \EndWhile
            \State Initialize $\delta_x$ and $\delta_f$ 
            \While{not converged} \Comment{Adversarial Perturbation Generation}
                \State Apply $\delta_f$ to $f_d$ and get $\hat{f}_d$
                \State $\hat h_d^{aux}, h_d^{aux} \gets \hat f_d(x_d^{aux}+\delta_x), f_d(x_d^{aux})$
                \State $\mathcal{L}_{dist}\gets {||\hat h_d^{aux}-h_d^{aux}||}_2$
                \State $\mathcal{L}_{mis}\gets -log(1-p(y_d^{aux}|g_s(\hat h_d^{aux})))$
                \State $\mathcal{L}\gets \lambda_1 \mathcal{L}_{dist}+\lambda_2 \mathcal{L}_{mis}$
                \State $\delta_x,\delta_f\gets \mathit{Backward}(\mathcal{L})$
            \EndWhile
            \State Apply $\delta_f$ to $f_d$ and get $\hat{f}_d$
            \State \textbf{return} $\delta_x,\hat{f}_d$
        \EndProcedure
        \Procedure{OnlineAttack}{$x_d,\delta_x,f_d,\hat{f}_d$}
            \For{each sample $x^i_d$ in $x_d$} 
                \If{\data launch \attack}
                        \State Send $\hat{f}_d(x_d^{i}+\delta_x)$ to \coord  \Comment{Perform attack}
                    \Else
                        \State Send $f_d(x_d^{i})$ to \coord
                \EndIf
            \EndFor
        \EndProcedure
    \end{algorithmic}
\end{algorithm}

\myparagraph{Surrogate Model Training}.
\data trains a surrogate top model $g_s$ using its auxiliary labeled set $\{x_d^{aux}, y_d^{aux}\}$ to obtain a differentiable mapping from the \data's representation to labels (Lines 2-6 in Alg.~\ref{alg:attack}).
Importantly, the adversary does not need to know the architecture of $g$.
The key requirement is that $g_s$ captures a coarse decision boundary to approximate the gradient.


\myparagraph{Adversarial Noise Generation.} 
Based on the trained $g_s$, we generate the adversarial noises $\delta_x$ and $\delta_f$ to construct the adversarial representation (Lines 7-17 in Alg.~\ref{alg:attack}).
The adversarial noises are initialized randomly, and both $\delta_x$ and $\hat{f}_d$ are updated iteratively by minimizing the distance between the benign representation $h_d^{aux}$ and the adversarial representation $\hat h_d^{aux}$, while simultaneously maximizing the distance between the true prediction $y_d^{aux}$ and the attacked prediction $\hat y_d^{aux}$ (Lines 12-14 in Alg.~\ref{alg:attack}).
In each iteration, we sample a batch of surrogate auxiliary samples and compute the gradient with respect to the loss function.
After the pre-defined number of iterations, the adversarial noises are generated and fixed.
Since the gradient backpropagation updates $\hat{f}_d$, we do not need to add explicit noises to $\hat{f}_d$ during the attack.
\data obtains the trained $\delta_x$ and $\hat{f}_d$ to launch \attack in \infer.

\myparagraph{Launch Attack.}
\data selectively initiate \attack on a subset of test samples (Lines 18-24 in Alg.~\ref{alg:attack}).
For each selected sample, \data generates tampered $\hat{x}_d$ and $\hat{f}_d$ by using adversarial noises to the original data $x_d$ and model $f_d$.
The adversarial representations $\hat{h}_d$ exhibit negligible statistical difference from the true representation $h_d$, yet it still leads to incorrect predictions in the \task's inference.

\subsection{Evaluation Setup}
\label{subsec:attack_eval_setup}

\myparagraph{Testbed.}
We implement the prototype of \attack and evaluate the performance on a server with NVIDIA RTX 4070ti super, Intel 9700KF, 32GB of memory, PyTorch 2.1, and CUDA driver 12.0.

\myparagraph{Dataset, Models and Partitions.}
We conduct the experiments on four representative datasets: Bank Marketing (BM)~\cite{MoroCR14}, Credit Card Fraud Detection (CCFD)~\cite{YehL09a}, Medical-MNIST (MMNIST)~\cite{medicalmnist}, and CIFAR10~\cite{krizhevsky2009learning}.
The choice of datasets follows the common practice of prior VFL studies~\cite{HuangW023,LuoWXO21,naseri2023badvfl}.
Recall that in Sec.~\ref{subsec:vfl}, we consider one $\mathcal{P}_t$ and one $\mathcal{P}_d$. Thus, we divide the data into two splits along the feature dimension. 
Model architectures and partitioning strategies are consistent with prior work\cite{BaiCZ0WG23,PangYSW23}.
Detailed configurations are provided in Tbl.~\ref{tab:dataset} and Tbl.~\ref{tab:model_split} in Appx.~\ref{app:eval_setup}.

\myparagraph{Attack and Defense Setting.}
In our implementation, we instantiate $g_s$ as a two-layer FCNN for simplicity and train $ g_s$ on $10\%$ of the training labels, following prior practice~\cite{ChoHYLBP24}.
We use the malicious tampering rate $K$ as the percentage of malicious samples injected by \attack out of all samples.
Currently, there are no specific defense mechanisms designed for \attack.
We choose three state-of-the-art poisoning defenses in VFL to evaluate the effectiveness of \attack: VFedAD~\cite{LaiW0LZ23}, P-GAN~\cite{ChenZLGW24}, and VFLIP~\cite{ChoHYLBP24}. 

\myparagraph{Metrics.}
We follow prior work and use five metrics to evaluate the attack performance~\cite{ChoHYLBP24,KumariRFJS23,LiD24,NguyenRCYMFMMMZ22}: Accuracy, Attack Success Rate (ASR), Positive Predictive Value (PPV), True Positive Rate (TPR), and Negative Predictive Value (NPV).
We use accuracy and ASR to evaluate the attack performance and use the other three metrics to evaluate the defense effectiveness of prior defenses.
These evaluation metrics are described in detail in Appx.~\ref{app:eval_metrics}.
For PPV, TPR, and NPV, a higher value indicates a better defense performance.

\subsection{Evaluation}
\label{subsec:attack_evaluation}

\myparagraph{Attack Performance.}
We first evaluate the attack performance of \attack on four datasets in Tbl.~\ref{tab:attack_performance}.
We assume all the samples from \data are attacked to only focus on the attack influence (i.e., $K=100\%$).
Compared to regular inference without attack, \attack can reduce the \task's accuracy by an average of $34.49\%$.
For the binary classification tasks (BM and CCFD), \attack can reduce accuracy to nearly $50\%$ (random guessing).
This result shows that \attack is able to significantly reduce the joint inference performance during \infer.
In Appx.~\ref{app:attack_eval}, we evaluate the attack performance under different numbers of $g_s$'s layers and various $K$.
The results show that the accuracy gradually decreases with the increase of $K$.
The attack is effective when choosing other settings of $g_s$.
When the depth of $g_s$ matches the depth of \task's $g$, \attack can achieve the largest accuracy reduction and the highest ASR.

\begin{table}[!t]
\small
\caption{Attack performance of \attack. \attack can reduce the accuracy of VFL by an average of 34.49\%.
}
\begin{center}
\adjustbox{max width=0.95\linewidth}{
\begin{tabular}{cccc}
\toprule
\textbf{Dataset} & \textbf{Normal Accuracy} & \textbf{Attack Accuracy} & \textbf{ASR} \\ \midrule
BM      & 74.97\% & 54.22\% & 28.92\% \\ \midrule
CCFD    & 97.97\% & 48.64\% & 51.36\% \\ \midrule
MMNIST  & 97.26\% & 61.70\% & 36.24\% \\ \midrule
CIFAR10 & 66.99\% & 34.67\% & 36.82\% \\ \bottomrule
\end{tabular}
}
\label{tab:attack_performance}
\end{center}
\vspace{-4mm}
\end{table}

\myparagraph{Resistant to Existing Defenses.}
We further evaluate existing defenses against \attack in Tbl.~\ref{tab:existing_defense}.
We set the number of attacked inferences to be equal to the number of benign instances ($K=50\%$).
The three defenses achieve an average PPV and NPV of 52.54\% and 53.46\%, respectively.
The defense effectiveness is close to random guessing.
This suggests that existing defenses struggle to differentiate between attacked and benign inferences.
The average TPR is 29.28\%, indicating that these defenses can only identify a small fraction of malicious inferences.
A large number of attack inferences are undetected by the defenses.


\begin{table}[!t]
\small
\caption{Detection results of existing defenses against \attack.
Existing defenses fail to detect \attack effectively, achieving only 52.53\%, 29.28\%, and 53.46\% on average for PPV, TPR, and NPV, respectively.
}
\begin{center}
\adjustbox{max width=0.95\linewidth}{
\begin{tabular}{cccccc}
\toprule
\textbf{Defense} & \textbf{Metrics} & \textbf{BM} & \textbf{CCFD} & \textbf{MMNIST} & \textbf{CIFAR10} \\ \midrule

\multirow{3}{*}{\textbf{VFedAD}} & PPV & 56.32\% & 58.86\% & 52.73\% & 53.44\% \\
                                 & TPR & 36.87\% & 27.20\% & 25.54\% & 30.97\% \\
                                 & NPV & 57.63\% & 53.47\% & 53.12\% & 54.76\% \\ \midrule
\multirow{3}{*}{\textbf{P-GAN}}  & PPV & 44.24\% & 53.85\% & 53.10\% & 44.72\% \\
                                 & TPR & 30.67\% & 22.79\% & 26.28\% & 25.38\% \\
                                 & NPV & 51.86\% & 51.17\% & 55.68\% & 50.34\% \\ \midrule
\multirow{3}{*}{\textbf{VFLIP}}  & PPV & 50.13\% & 54.32\% & 52.56\% & 56.12\% \\
                                 & TPR & 34.20\% & 30.33\% & 28.46\% & 32.28\% \\
                                 & NPV & 55.25\% & 50.45\% & 53.65\% & 54.17\% \\ \bottomrule 
\end{tabular}
}
\label{tab:existing_defense}
\end{center}
\end{table}

\myparagraph{Summary.}
Based on the evaluation results, we can conclude that \data can use \attack to degrade the performance of \task's inference accuracy effectively.
Besides, the existing defense in VFL can not detect the attacked inferences due to the stealthiness of \attack.

\section{\method Defense}
\label{sec:defense}
The success of \attack attack is because no VFL schemes offer efficient validation mechanisms to audit \data's behavior in \infer. 
In this section, we will introduce our defense, \underline{Ve}rtical \underline{F}ederated \underline{I}nference \underline{A}udit framework (\method).
We first discuss the insight of our defense (Sec.~\ref{subsec:insight}), and introduce three challenges of our insight (Sec.~\ref{subsec:challenge}).
Then, we introduce three key components of \method: privacy-aware training (Sec.~\ref{subsec:training}), runtime authenticity validation (Sec.~\ref{subsec:authenticity}), and efficiency-aware computation schedule (Sec.~\ref{subsec:efficiency_schedule}).
Finally, we illustrate the workflow of \method in Sec.~\ref{subsec:workflow}.

\subsection{Insight}
\label{subsec:insight}

The central challenge in securing the joint inference is the \textbf{privacy-auditing contradiction}: the inherent conflict between preserving \data's data privacy and enabling effective auditing inference.
In VFL frameworks, \task cannot censor \data's inference process due to privacy restrictions. 
As revealed by our attack, \task cannot directly validate the authenticity of $f_d$ and $x_d$ in \data, but \data can easily tamper with them to degrade inference performance.

Our insight to solve the privacy-auditing contradiction is to leverage the TEE-COO framework to assist \task to audit \data's inference behaviors.
The security features of TEE allow \task to run a secure validation program on \data.
Specifically, \data needs to load the data and model into the TEE to run the TEE-COO inference program and generate trusted inference results as evidence.
Meanwhile, \data performs inference on its local untrusted accelerator (e.g., a GPU). 
\task can use the trusted inference results to validate the untrusted inference results and audit whether \data's inference is executed as expected.
In this way, if \data tampers with the inference process, \task can detect inconsistent inference and stop inferring on $g$ and reclaim the inference fee.
The trusted pipeline provided by \data's TEE and remote \coord ensures the integrity of the inference process.
For remote outsourcing, we follow the motivation of TEE-COO~\cite{abs-1912-03485,mo2020darknetz} and split \data's bottom model $f_d$ into two parts along the depth dimension: a shallow bottom model $f_d^{sm}$ running on \data's TEE and a deep bottom model $f_d^{dm}$ running in the \coord.
The insight of this design is that the shallow layers of the bottom model are more privacy-sensitive and contain more private information of \data's data~\cite{abs-1912-03485}.
TEE-COO mechanism ensures that \data's inputs remain confined within its secure domain, while \coord is only permitted to receive the intermediate inference result $z_d=f_d^{sm}(x_d)$.

\subsection{Technical Challenges}
\label{subsec:challenge}


Using TEE-COO trusted inference to audit \data's untrusted inference is a straightforward solution but faces several technical challenges.

\myparagraph{C1: Privacy Leakage Measurement.}
The first challenge is mitigating \data's privacy leakage during inference outsourcing.
One primary motivation of VFL is to protect the data privacy of \data. 
However, introducing the TEE-COO framework requires deploying partial layers of $f_d$ to the \coord.
Prior work has proved that \coord may recover $x_d$ through $z_d$, which is called model inversion attacks~\cite{HeZL19,YinZZLYCH23}.
Prior TEE-COO framework used a proof-of-concept attack to measure privacy leakage.
For example, DarkneTZ uses a gradient-based membership inference attack to calculate how much privacy each layer leaks and determines privacy protection configuration~\cite{mo2020darknetz}.
However, proof-of-concept attacks can not provide a rigorous privacy guarantee.
A strong attack may break the optimal configuration defined by old attacks.
Thus, we need to rigorously measure the additional privacy leakage introduced by TEE-COO to prevent unknown stronger attacks.



\myparagraph{C2: Runtime Subject Authenticity.}
The second challenge is checking the authenticity of the data and model used for TEE-COO collaborative inference.
The confidentiality of TEE ensures the integrity of enclave computation.
However, the TEE-COO inference program is executed at \data.
\data can still load tampered $\hat{x}_d$ and $\hat{f}_d$ into the TEE at runtime.
This raises the issue that the authenticity of the inference subjects cannot be validated, and the trustworthiness of the TEE-COO collaborative inference is not guaranteed.

\myparagraph{C3: Validation Efficiency.}
The third challenge is how to efficiently audit the untrusted inference during \infer with a large volume of queries.
As mentioned in Sec.~\ref{subsec:vfl}, each inference task on VFL requires performing inference on many samples within a short time frame.
According to the analysis in Sec.~\ref{subsec:attack_evaluation}, \data can use \attack to launch widespread attacks on local inference.
Although the TEE-COO can improve the speed of trusted inference, it is still slower than the untrusted inference on the GPU alone.
Deploying inference auditing for every sample would reduce the system's throughput, leading to a long waiting time for inference services.
Thus, we need to design an efficient validation method that can quickly assist \task to audit whether \data has launched \attack.

\subsection{Privacy-Aware Training}
\label{subsec:training}


To address \textbf{C1}, we propose a privacy-aware training solution for \train based on information theory~\cite{AlemiFD017,belghazi18a,ChengHDLGC20}.
We partition the bottom model $f_d$ of \data into two sub-models: $f_d^{sm}$ and $f_d^{dm}$.
Then we use mutual information to quantify the extent to which intermediate representations $z_d=f_d^{sm}(x_d)$ retain private information about $x_d$~\cite{NoorbakhshZHW24,TangDDDHG0L24,Tan00LG024}.
During \train, we learn a perturbation $\sigma$ that is added to the representations of $z_d$: $\hat z_d=z_d+\sigma$.
The goal is to minimize mutual information between the perturbed representations $\hat z_d$ and  $x_d$.
As a result, when \data transmits $\hat z_d$ to \coord during \infer, the privacy exposure in the shared representation is effectively bounded.
We use mutual information neural estimation~\cite{belghazi18a}, which employs a DNN model to estimate mutual information between two variables $a$ and $b$ as follows.
\begin{equation}
    I(a;b)=\mathbb{E}_{p(a,b)}[V(a,b)]-\log\mathbb{E}_{p(a)p(b)}[e^{V(a,b)}].
\end{equation}
$V$ is a DNN model that approximates the ratio of joint distribution to the product of the marginals.
$p(a,b)$ denotes the joint distribution of variables $a$ and $b$, while $p(a)$ and $p(b)$ denote the marginal distribution of $a$ and $b$.
Specifically, in the context of VFL, the two variables are $x_d$ and $\hat z_d$.
To minimize the privacy in $\hat z_d$, we incorporate $\mathcal{L}_{\mathit{mi}}=I(x_d;\hat z_d)$ into standard VFL loss function:
\begin{equation}
    \mathcal{L}=\mathcal{L}_{\mathit{task}}+\lambda \mathcal{L}_{\mathit{mi}}
\end{equation}
where $\mathcal{L}_{\mathit{task}}$ is the main task loss in VFL and $\lambda$ is a hyperparameter that balances the different loss components.
After \train, \data sends the trained $f_d^{sm}$ and $\sigma$ to local TEE and sends $f_d^{dm}$ to \coord.

\myparagraph{Theoretical Analysis.}
We further analyze the theoretical guarantee we can provide against inference-time model inversion attacks~\cite{HeZL19,YinZZLYCH23,abs-2406-12588,LuoWXO21}.
Model inversion is a privacy leakage attack in cloud-edge collaborative inference, where the cloud server attempts to reconstruct raw data from the intermediate representations provided by the edge.
Let $\mathcal{A}$ be an arbitrary attack model that \coord can use to infer private inputs $x_d$ from $\hat z_d$.
\begin{lemma}
    Let $h(\cdot)$ be the differential entropy function, the mutual information $I(x_d;\hat z_d)$ can be expressed as:
    \begin{equation}
        I(x_d;\hat z_d)=h(x_d)-h(x_d|\hat z_d)=h(\hat z_d)-h(\hat z_d|x_d)=h(\hat z_d)
    \end{equation}
\end{lemma}
Since the mapping from $x_d$ to $\hat{z}_d$ is fixed after training, there is no uncertainty, i.e., $h(\hat z_d|x_d)=0$.
$H(\hat{z}_d)$ thus quantifies the total information retained in $\hat{z}_d$.
While deep neural networks tend to compress inputs by preserving only task-relevant features, the shallower structure of $f_d^{sm}$ retains more input information, resulting in higher entropy.
As a result, mutual information becomes a key factor in characterizing model inversion.
We formalize the relation between mutual information and inversion error through the following theorem (see proof in Appx.~\ref{app:proof}):
\begin{theorem}
\label{theorem:bound}
    (Lower bound for inversion error).
    Let $x_d \in \mathbb{R}^m$, the lower bound of inversion error can be formalized as:
    \begin{equation}
    \label{eq:lower_bound}
        \mathbb{E}(||x_d-\mathcal{A}(\hat z_d)||_q/m)\ge \frac{e^\frac{2}{m}h(x_d)}{2\pi e}e^{-\frac{2}{m}I(x_d;\hat z_d)}
    \end{equation}
    where $m$ is the feature dimension of $x_d$ and $q$ is the $q$-norm.
\end{theorem}
This means that the inversion error is exponentially negatively correlated with mutual information.
A lower mutual information $I(x_d;\hat{z}_d)$ implies a higher lower bound on the inversion error, making it more difficult for \coord to reconstruct the original data $x_d$.
Privacy-aware training continuously reduces mutual information by adding $\mathcal{L}_{mi}$ to VFL training loss, thereby reducing the risk of data leakage.


\subsection{Runtime Authenticity Validation}
\label{subsec:authenticity}

Statistical-based detection techniques that are widely used in data poisoning defenses are susceptible to false positives and false negatives~\cite{stokes2021preventing}.
In real-world deployments, authentication and provenance-based mechanisms provide a more robust and reliable defense for ML systems~\cite{stokes2021preventing,baracaldo2017mitigating,guo2021modelshield}.
Inspired by this, we propose a hash consistency validation mechanism to check the authenticity of $x_d$ and $f_d$ to address \textbf{C2}.
The validation process runs within the TEE, thus \data cannot interfere with the process.
The mechanism consists of two components: data validation and model validation.

\myparagraph{Data Validation.}
This mechanism ensures data authenticity by comparing the hash value during \infer with the stored counterpart.
Specifically, before \preprocess, TEE computes the hash signature of the dataset and encrypts it to prevent \data from accessing or modifying them.
During \infer, whenever TEE loads data, \method recomputes the hash and compares it against the stored value.
A match confirms the data authenticity.
Note that \data cannot tamper with the data before TEE's hash precomputation, as this step occurs before \preprocess~\cite{stokes2021preventing}.
At this stage, \data lacks knowledge of which samples will be used for training or inference.
Any tampering before \preprocess may corrupt the training dataset, and potentially degrade model accuracy.
If the training accuracy fails to meet the required threshold, \task will refuse to collaborate with \data.

\myparagraph{Model Validation.}
\data used privacy-aware training to partition $f_d$ into $f_d^{sm}$ (loaded into TEE) and $f_d^{dm}$ (outsourced to \coord). 
The purpose of model validation is to ensure that the model received by TEE and \coord is the authentic model trained by VFL.
\task secretly sends the IDs of validation samples to TEE.
\method loads the corresponding samples into TEE for data validation and subsequently initiates the TEE-COO collaborative inference. 
\task validates that the model accuracy meets the required performance threshold, after which \method stores a cryptographic hash of the model weights.
Similar to data validation, this hash is signed and encrypted by TEE, preventing \data from tampering with it.
Notably, the performance validation occurs only once, as \task does not require further updates once the model achieves the desired performance.
For each inference query, \method loads the model weights into TEE, computes their hash value, and validates its consistency with the stored post-training validation hash.
If a discrepancy is detected, indicating that \data has modified the model weights, \method issues a warning to \task.



\subsection{Efficiency-Aware Computation Schedule}
\label{subsec:efficiency_schedule}
To address \textbf{C3}, our validation mechanism must be efficient and accurate enough for \infer.
Thus, we design a confidential random sampling validation mechanism that allows \task to use TEE-COO trusted inference to validate \data's untrusted inference (as shown in Appx.~\ref{app:crsv}).
This mechanism allows trusted and untrusted inferences to run in parallel and requires fewer trusted inferences to detect malicious behavior of \data.

Suppose $N$ samples need to be inferred in an inference query.
$K$ is the ratio of $N$ samples that are tampered by \data, and $W$ is the ratio of samples that \task can secretly validate in TEE. 
Our key insight is that, for efficiency reasons, \task cannot audit every individual inference from \data during \infer.
A more practical and effective approach is for \method to secretly validate a designated set of $WN$ samples ($W \cdot 100\%$ of the dataset); if any inference instance is detected to be malicious within this set, \task can reasonably conclude that \data is acting maliciously.
We can formulate as follows to give a theoretical Defense Success Rate (DSR): 
\begin{equation}
\label{eq:prob}
DSR=1-\binom{N-\mathit{K}N}{WN}/{\binom{N}{WN}}
\end{equation}
$\binom{N}{WN}$ refers to the process of randomly sampling $WN$ instances from $N$ total inferences for validation.
$\binom{N-\mathit{K}N}{WN}$ denotes a sampling of $WN$ validation instances that does not include any tampered inferences.
The motivation of this mechanism is that \task can secretly send the target IDs to the enclave via TEE attestation for inference validation.
Therefore, the sample IDs inferred within the enclave remain a complete black box to \data.

\myparagraph{Efficiency Optimization.}
\label{subsec:eo}
To reduce latency, we aim to balance the time between trusted and untrusted inference.
The system will not incur additional latency if the trusted and untrusted inference times are equal. 
Let $T_{un}$ and $T_{tr}$ represent the inference time for \data on $N$ samples using an untrusted accelerator and TEE-COO trusted inference, respectively.
The optimal sampling ratio $W_{*}$ is given by:
\begin{equation}
\label{eq:osr}
    W_{*}=\myargmin_W|T_{un}-W T_{tr}|
\end{equation}
The objective is to find the optimal $W_*$ such that the total time to complete $WN$ trusted inferences (via TEE-COO; $W\times 100\%$ of the $N$ samples) equals the time required to complete $N$ untrusted samples (on GPU).
\task can require \data to perform trusted inference on validation samples in the first model validation phase (Sec.~\ref{subsec:authenticity}) and compute $W_*$ with historical training information.
Specifically, \method provides an efficiency-optimal inference validation strategy:
\begin{theorem}
\label{theorem}
    (Optimal sampling auditing guarantee).
    Let an inference query consist of $N$ samples. Given an optimal sampling ratio $W_*$ and an expected defense success rate (e.g., $99.99\%$), \method guarantees that if the malicious tampering rate $K$ exceeds a threshold, then the probability of detecting at least one instance of malicious inference is not less than the expected defense success rate, and the auditing process does not introduce additional system latency.
\end{theorem}
The goal of our mechanism is to ensure that, once the number of attacked inferences surpasses a certain threshold, the task can detect malicious behavior from \data with high probability.
Notably, the goal is not to pinpoint individual malicious inferences but to reliably detect the presence of an attack.


%

\myparagraph{Pipeline Acceleration.}
\label{subsec:pipeline}
\method's trusted inference relies on collaborative inference between the \data's TEE and \coord's GPU.
\coord needs to wait until the local enclave completes inference on all samples.
In scenarios with a large number of model parameters, the computing power of TEE will become a bottleneck, causing \coord to be idle for too long.
Thus we use a block inference pipeline execution to accelerate collaborative inference and improve the system efficiency.
Trusted inference consists of three operations $o=\{o_{TEE},o_{comm},o_{coo}\}$, which correspond to the TEE computation, the TEE-COO communication, and the \coord computation, respectively.
During the first model validation, \method evaluates the TEE-COO inference latency $T_o$ of three operations for the determination of the optimal number of blocks $B_*$.
To avoid complex combinations, we set $o_{comm}$ to a constant~\cite{Jiang0C24}.
We transform this process into an optimization problem, specifically targeting the minimization of the completion time $\beta^{end}_{O,B}$ for the final operation $O$ of the last block $B$:
\begin{align}
B_{*} &= \myargmin_{B} \beta^{end}_{O,B} \label{eq:optimal_block} \\ 
\text{s.t.} \quad  \beta^{end}_{o,b} &= \beta^{start}_{o,b} + T_o, \quad B + f_d^{sm} \leq EPC \label{eq:epc} \\
{order}_{o,b} &= \mathbb{I}\,({o\ne0}) \cdot \beta^{end}_{o-1,b} \label{eq:order} \\
{resource}_{o,b} &= \mathbb{I}\,{((o,b) \neq (0,0))} \cdot \beta^{end}_{o,b-1} \label{eq:resource}
\end{align}
%
where $b\le |B|, o\le |O|$.
Eq.~\ref{eq:epc} ensures that the size of the block and $f_d^{sm}$ does not exceed the Enclave Page Cache (EPC) size.
Eq.~\ref{eq:order} ensures that each block executes all operations sequentially.
Eq.~\ref{eq:resource} stipulates that a given resource can be allocated to only one block at any given time.
The optimal configuration $B^{*}$ can be determined by a limited number of enumerations.

\begin{figure}[!t]
\centerline{\includegraphics[width=1\linewidth]{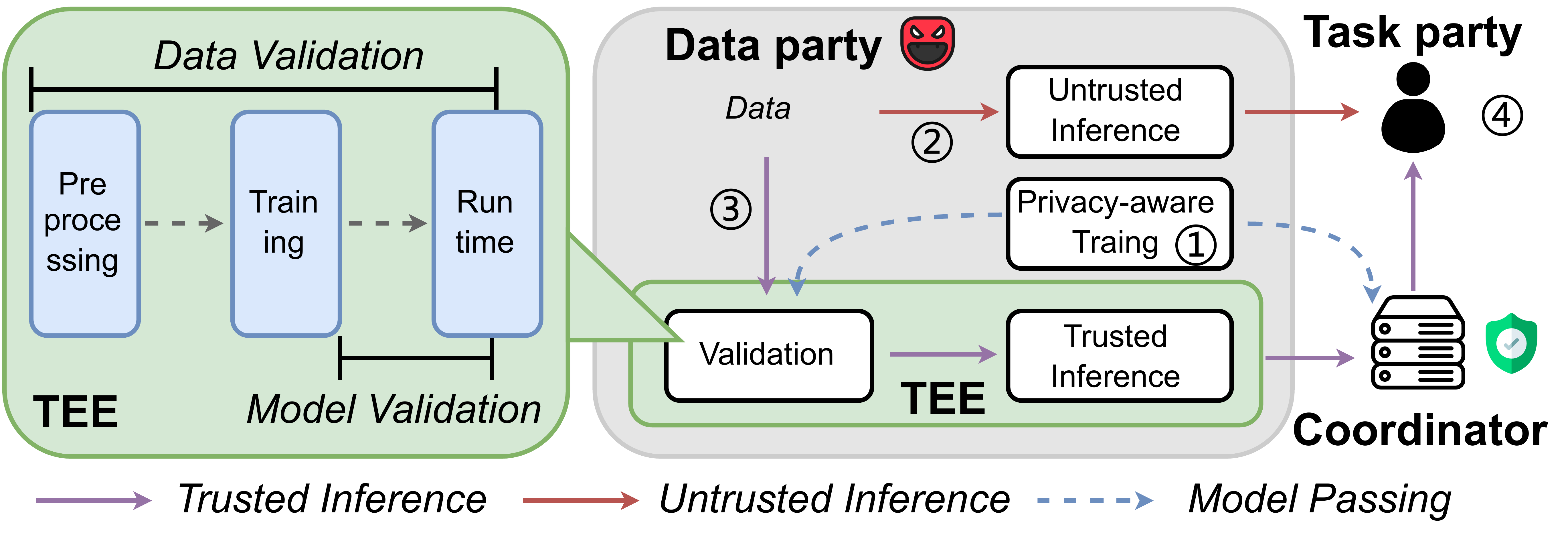}}
\caption{Workflow of \method.
\ding{172} \data adopts privacy-aware training to produce $f_d^{sm}$ and $\sigma$, which are sent to TEE, and $f_d^{dm}$, which is sent to \coord.
\ding{173} \data performs untrusted inference and transmits the results to \task.
\ding{174} TEE performs model and data validation, and conducts trusted collaborative inference with \coord, and sends results to \task.
\ding{175} \task uses the trusted inference results to audit the authenticity of the untrusted inference.
}
\vspace{-1.5em}
\Description{}
\label{fig:overview}
\end{figure}

\subsection{Workflow}
\label{subsec:workflow}

Combining the three technical designs of \method, Fig.~\ref{fig:overview} shows the \method's workflow.
The workflow consists of four steps:

\noindent \textbf{\ding{172} Privacy-Aware Training.} 
During \train, \data partitions the model $f_d$ into $f_d^{sm}$ and $f_d^{dm}$ to facilitate privacy-aware training solution (details in Sec.~\ref{subsec:training}), ensuring that the output of $f_d^{sm}$ preserves minimal private information while maintaining the overall VFL performance.
Upon completion of \train, \data transfers $f_d^{sm}$ and $\sigma$ to local TEE and $f_d^{dm}$ to \coord for validation and inference.

\noindent \textbf{\ding{173} Untrusted Inference.}
\task sends the IDs of $N$ targeted samples for inference queries to \data.
\data performs local inference using the corresponding $x_d$ and $f_d$.
During this process, \data may launch \attack on partial inferences to produce abnormal results.
Upon completion, \data returns the $N$ untrusted inference results to \task.

\noindent \textbf{\ding{174} Trusted Inference.}
\task uses the confidential random sampling validation (details in Sec.~\ref{subsec:efficiency_schedule}) to select a validation subset from $N$ inference requests and sends the IDs to \data's TEE.
The TEE first runs the validation code, loading $x_d^v$ and $f_d^{sm}$ to perform data validation and model validation.
Data validation ensures that $x_d^v$ at runtime is consistent with its stored state before \preprocess.
Model validation ensures that $f_d^{sm}$ and $f_d^{dm}$ at runtime are consistent with their state at the end of \train.
After both validation passes (details in Sec.~\ref{subsec:authenticity}), the TEE executes the inference code and sends the trusted inference results to the \coord.
\coord performs the final inference using $f_d^{dm}$ to complete the TEE-COO trusted collaborative inference.

\noindent \textbf{\ding{175} Inference Consistency Validation.}
Untrusted (\ding{173}) and trusted (\ding{174}) inference run in parallel.
\task audits whether the \data's inference program was executed as expected by checking the consistency of the inference results.

\section{Defense Evaluation}
This section evaluates the performance of \method.
We first describe the setup (Sec.\ref{subsec:defense_setup}), followed by the defense effectiveness (Sec.\ref{subsec:defense_eval}), privacy protection results (Sec.\ref{subsec:privacy_eval}), system efficiency on real hardware (Sec.\ref{subsec:system_efficiency}) and scalability with increasing numbers of \data.

\subsection{Evaluation Setup}
\label{subsec:defense_setup}

\myparagraph{Implementation.} 
We implemented \method on the same platform as Sec.~\ref{subsec:attack_eval_setup}.
The TEE module is developed in C++ using the Eigen 3.2.10 library and compiled by Intel SGX SDK 2.23 (Intel-supported TEE) and GCC 11.4.0.

\myparagraph{Configurations.}
We validate \method on the same four datasets and models as Sec.~\ref{subsec:attack_eval_setup}.
To additionally assess the efficiency, we further evaluate it on a large-scale KDD Cup 1999 dataset (KDD-CUP)~\cite{TavallaeeBLG09}.
Common VFL datasets (e.g. BM, CCFD, MMNIST, and CIFAR10 datasets) typically contain tens of thousands of samples, whereas the KDD-CUP dataset reaches a million-scale magnitude (10 to 100 times larger).
KDD-CUP dataset is used to evaluate the efficiency (Sec.~\ref{subsec:system_efficiency}) and scalability (Sec.~\ref{subsec:scalability}) of \method.
The VFL partitioning scheme is shown in Tbl.~\ref{tab:dataset} and Tbl.~\ref{tab:model_split}. 
We follow prior literature to set the communication cost between \data and \coord as 1 second~\cite{Jiang0C24}.


\myparagraph{Metrics.} 
We used the three metrics mentioned in Sec.~\ref{subsec:attack_eval_setup} (PPV, TPR, and NPV), along with four new metrics:
\begin{itemize}[leftmargin=*]
    \item \textit{Detection Success Rate (DSR)}:
    The probability of detecting a malicious inference in \method's random sampling validation (Eq.~\ref{eq:prob}).
    \item \textit{Speedup Ratio}:
    Speed ratio is the inference latency improvement of \method (w/o pipeline acceleration) over \method.
    A higher value indicates lower overall system latency, demonstrating the efficiency gains achieved through pipeline optimization.
\end{itemize}
Following prior work~\cite{LiuF00024,abs-2404-15821,WangBSS04}, we evaluate the similarity between original and inversed data using Hitting Rate (HR) for tabular data (BM and CCFD) and Structural Similarity Index Measure (SSIM) for image data (MMNIST and CIFAR10).
We adopt safety thresholds of 0.09 for HR and 0.3 for SSIM~\cite{abs-2404-15821,HeZL19}, below which inversion is considered unsuccessful (details in Appx.~\ref{app:defense_metrics}).

\myparagraph{Attack Setup.} 
We use two types of attacks to evaluate \method.
The first type is proposed tampering attack \attack.
The setting is consistent with Sec.~\ref{subsec:attack_eval_setup}.
We will fully verify the detection effect of \method on \attack in the experiment.
The second type is model inversion attacks.
We use these attacks to evaluate our privacy-aware training mechanism and how much \method can protect the input privacy of \data.
Specifically, we select four state-of-the-art inversion attacks (details in Appx.~\ref{app:moia}): query-free model inversion~\cite{HeZL19}, Genver~\cite{YinZZLYCH23}, UIFV~\cite{abs-2406-12588}, and FIA~\cite{LuoWXO21}.
The settings of these attacks are consistent with the original paper.


\subsection{Theoretical vs Empirical Validation}
\label{subsec:main_eval}
\method guarantees a high detection success rate DSR (e.g., $99.99\%$) when the malicious tampering rate $K$ exceeds a certain threshold.
Eq.~\ref{eq:prob} defines this theoretical threshold for $K$.
This experiment aims to verify whether DSR approaches the target level when $K$ is near this theoretical threshold.

According to Sec.~\ref{subsec:efficiency_schedule}, \method can use the validation samples during the first model validation to determine the optimal $W_*$.
Under optimal $W_*$ (column $W_*$ in Tbl.~\ref{tab:theoretical}), using \method for inference auditing does not introduce additional latency to the online inference system.
The $W_*$ for BM and CCFD are higher than those for MMNIST and CIFAR10 due to the greater computational overhead of the convolutional layers compared to the linear layers.

According to Thm.~\ref{theorem}, $W_*$ can achieve the expected DSR (e.g. 90\% or 99.99\%) when the malicious tampering ratio $K$ exceeds a certain value.
Using Eq.~\ref{eq:prob}, we compute a theoretical lower bound for $K$.
In Tbl.~\ref{tab:theoretical}, we further validate whether the real $K$ aligns with the theoretical lower bound.
Column $T$ in Tbl.\ref{tab:theoretical} represents the theoretical lower bound derived from Eq.~\ref{eq:prob}, indicating that as long as $K$ exceeds this bound, \method can detect malicious inferences by \data with a DSR of 90\% or 99.99\%.
In our experiment, we iteratively decreased the lower bound (in steps of $1\%$) to determine whether the actual DSR met the expected DSR.
The results in Tbl.~\ref{tab:theoretical} demonstrate that \method guarantees that malicious behavior can be detected with a probability of 99.99\% when $K$ exceeds an average threshold of 5.4\% in practice.
Furthermore, as shown in Tbl.~\ref{tab:theoretical}, the real values $K$ (Column $R$) across all four datasets are consistently lower than the theoretical lower bound (Column $T$).
This indicates that, in practical scenarios, \method can achieve the desired DSR for the \task with a smaller $K$ than theoretically required.

\begin{table}[!t]
\footnotesize
\caption{
Comparison of theoretical (T) and real (R) $K$ at $W_*$ to achieve the target DSR.
Under $W_*$, \method can detect \attack with 99.99\% probability when the proportion of malicious inferences $K$ exceeds 5.4\% on average.
}
\begin{center}
\adjustbox{max width=0.95\linewidth}{
\begin{tabular}{ccccccc}
\toprule
\multirow{2}{*}{\textbf{Dataset}} & \multirow{2}{*}{\textbf{N}} & \multirow{2}{*}{\textbf{$W    _*$}} & \multicolumn{2}{c}{\textbf{DSR=90\%}} & \multicolumn{2}{c}{\textbf{DSR=99.99\%}} \\ \cmidrule(lr){4-5} \cmidrule(lr){6-7}
        &        & (\%)  & \textbf{R(\%)} & \textbf{T(\%)} & \textbf{R(\%)} & \textbf{T(\%)} \\ \midrule
BM      & 21969  & 34.4  & 2.80e-2 & 2.80e-2 & 0.10    & 0.11    \\ \midrule
CCFD    & 170589 & 39.6  & 2.70e-2 & 2.94e-3 & 1.04e-2 & 1.12e-2 \\ \midrule
MMNIST  & 17687  & 0.42  & 2.84    & 3.05    & 10.40   & 11.69   \\ \midrule
CIFAR10 & 15000  & 0.35  & 4.27    & 4.33    & 11.21   & 16.15   \\ \bottomrule
\end{tabular}
}
\label{tab:theoretical}
\end{center}
\end{table}

\subsection{Defense Effectiveness}
\label{subsec:defense_eval}
This section evaluates the detection success rate DSR of \method against \attack with varying $K$ and $N$.
Additionally, we compare the performance of \method with that of existing defense mechanisms.

\begin{figure}[!t]
    \centering
    \begin{subfigure}{0.46\linewidth}
        \centering
        \includegraphics[width=\linewidth]{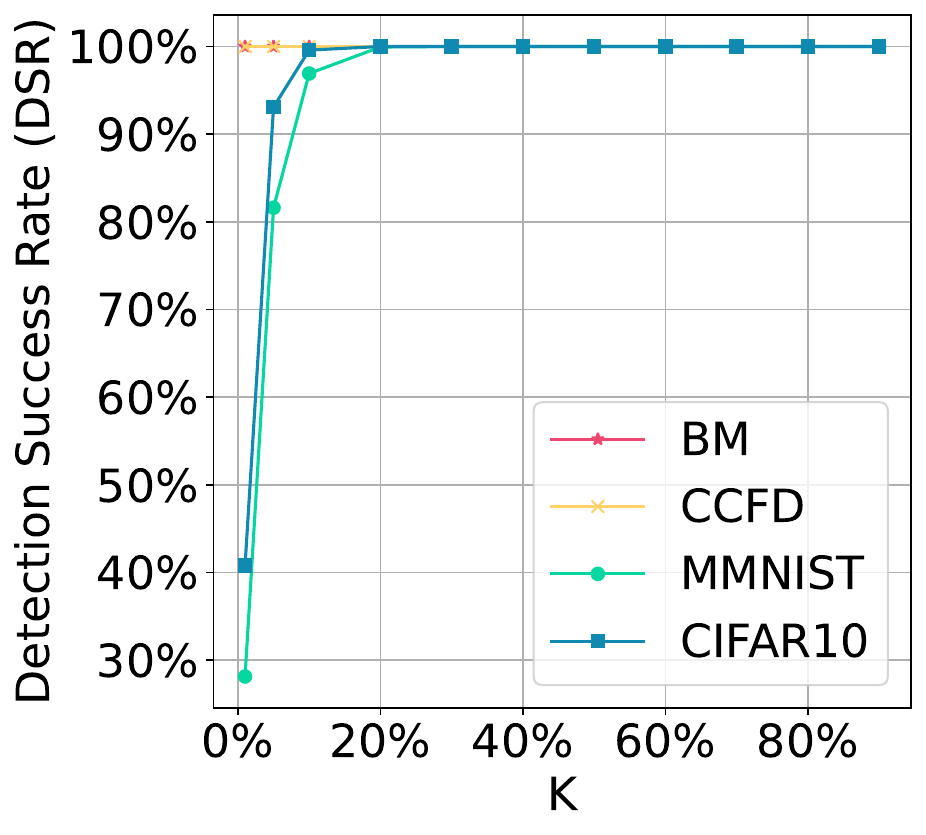}
    \end{subfigure}
    \hfill
    \begin{subfigure}{0.48\linewidth}
        \centering
        \includegraphics[width=\linewidth]{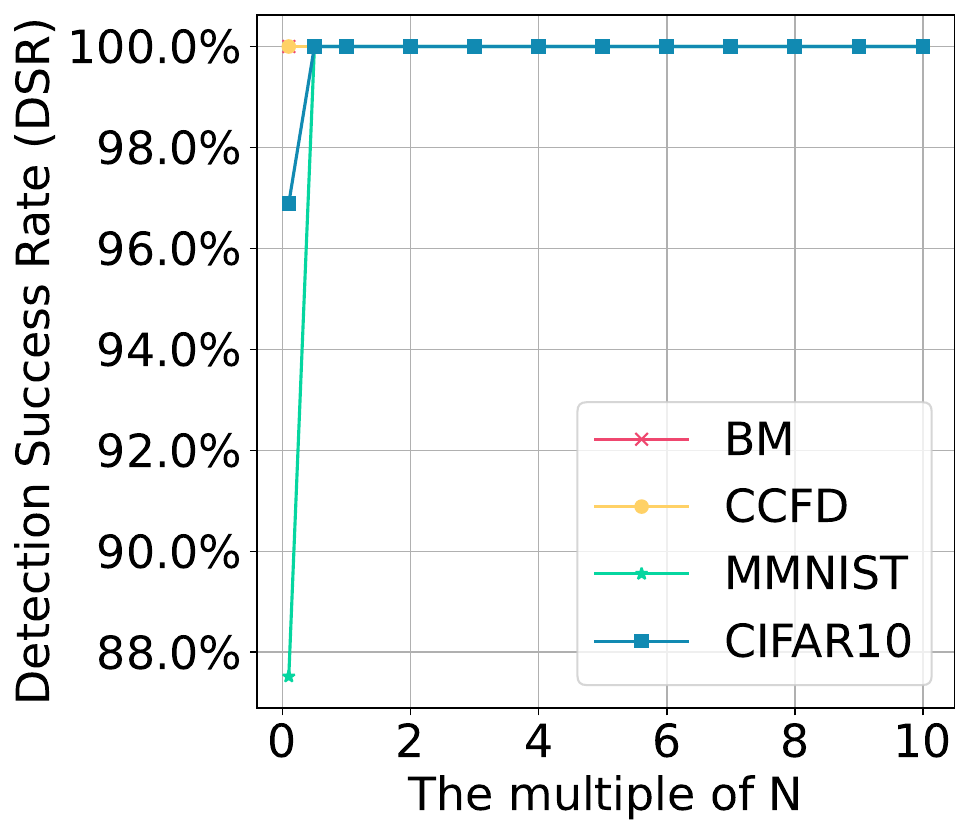}
    \end{subfigure}
    \caption{The change of detection success rate (DSR) with different $K$ (left) and $N$ (right).
    }
    \Description{}
    \label{fig:dsr}
\end{figure}

In Fig.~\ref{fig:dsr} (left), we gradually increase the $K\%$ ($1\%-90\%$) to evaluate the detection capability of \method against \attack under $W_*$.
On the BM and CCFD datasets, the \method consistently achieves a DSR of $100\%$.
On the MMNIST and CIFAR10 datasets, once $K\%$ exceeds $10\%$, \method can detect the malicious behavior with a probability greater than $96\%$.
As described in Sec.~\ref{subsec:attack_evaluation}, due to the limited knowledge of the \data during \infer, significantly degrading the \task's inference performance requires launching \attack on a large scale across many inferences.
However, this makes the \data's malicious behavior more easily detectable by \method, enabling the \task to hold the \data accountable for inference and impose penalties.

The design of \method targets \infer.
We incrementally increase the multiple of $N$ (using the Column $N$ in Tbl.~\ref{tab:theoretical} as the base) to validate the detection effectiveness of \method under high-throughput conditions.
As illustrated in Fig.~\ref{fig:dsr} (right), as the multiple of inference queries $N$ increases incrementally, \method achieves a $100\%$ DSR across all four datasets without incurring any additional system latency.
This also indicates that the design of \method is successful: it achieves an exceptionally high DSR ($100\%$) for malicious behavior of \data in high-throughput online inference systems with minimal validation costs.

\myparagraph{Comparison with existing defenses.}
As in Sec.~\ref{subsec:attack_evaluation}, we set $K$ to $50\%$.
We evaluate the defense performance of different mechanisms using PPV, TPR, and NPV on sampling samples covered by $W_*$.
The results in Tbl.~\ref{tab:vefia_defense} show that \method outperforms the other three mechanisms significantly in terms of defense performance.
The PPV and NPV of \method are both $100\%$, indicating that the inference consistency validation effectively captures subtle differences between trusted and untrusted inferences.
The TPR of $100\%$ indicates that the \method is able to detect all malicious inferences within the sampling samples that have been attacked by \attack.
Therefore, \data cannot bypass \method using \attack, and the \method will not produce a false positive.
The results in Appx.~\ref{app:detection_performance} indicate that, under different validation ratios $W$, \method can achieve a $100\%$ PPT, TPR, and NPV detection for \attack.
Note that Tbl.~\ref{tab:vefia_defense} is different from Tbl.\ref{tab:existing_defense}. Tbl.~\ref{tab:vefia_defense} presents the detection performance for samples covered by $W_*$, whereas Tbl.\ref{tab:existing_defense} evaluates detection across all test samples.

\begin{table}[!t]
\small
\caption{Comparison of defense performance on samples covered by $W_*$.
\method achieves 100\% PPV, TPR and NPV on sampled inferences.
}
\begin{center}
\adjustbox{max width=0.95\linewidth}{
\begin{tabular}{cccccc}
\toprule
\textbf{Defense} & \textbf{Metrics} & \textbf{BM} & \textbf{CCFD} & \textbf{MMNIST} & \textbf{CIFAR10} \\ \midrule

\multirow{3}{*}{\textbf{VFedAD}}  & PPV & 56.52\% & 58.98\% & 22.22\% & 46.67\% \\
                                  & TPR & 36.79\% & 27.12\% & 40.00\% & 29.17\% \\
                                  & NPV & 57.86\% & 53.20\% & 70.00\% & 54.05\% \\ \midrule
\multirow{3}{*}{\textbf{P-GAN}}   & PPV & 43.84\% & 53.14\% & 16.67\% & 38.46\% \\
                                  & TPR & 30.19\% & 23.77\% & 20.00\% & 20.83\% \\
                                  & NPV & 52.26\% & 51.44\% & 69.23\% & 51.28\% \\ \midrule
\multirow{3}{*}{\textbf{VFLIP}}   & PPV & 49.32\% & 53.39\% & 25.00\% & 50.00\% \\
                                  & TPR & 33.96\% & 29.81\% & 50.00\% & 25.00\% \\
                                  & NPV & 54.84\% & 50.77\% & 80.00\% & 55.00\% \\ \midrule
\multirow{2}{*}{\textbf{\method}} & PPV & 100.0\% & 100.0\% & 100.0\% & 100.0\% \\
                                  & TPR & 100.0\% & 100.0\% & 100.0\% & 100.0\% \\
\textbf{(Ours)}                   & NPV & 100.0\% & 100.0\% & 100.0\% & 100.0\% \\ 
\bottomrule
\end{tabular}
}
\label{tab:vefia_defense}
\end{center}
\end{table}

\subsection{Privacy Protection}
\label{subsec:privacy_eval} 
We use four privacy inference attacks to evaluate the effectiveness of the privacy-aware training mechanism (Sec.~\ref{subsec:training}) during outsourced inference.
HR is used to evaluate BM and CCFD, while SSIM is used to evaluate MMNIST and CIFAR-10.
As shown in Tbl.\ref{tab:eval_privacy}, \method maintains HR and SSIM below the safety thresholds of 0.09 and 0.3, respectively.
\method can reduce the similarity between original data and reconstructed data by an average of 72.9\%.
Furthermore, as shown in Tbl.~\ref{tab:accuracy}, the integration of privacy-aware training leads to only a 0.51\% average drop in VFL inference accuracy.
These results confirm that \method effectively protects \data's data privacy without compromising model performance.

\begin{table}[!t]
\caption{The protection performance of the privacy-aware training solution. For BM and CCFD, we report HR; for MMNIST and CIFAR10, we report SSIM. Our mechanism can reduce HR and SSIM by 72.9\% on average.
}
\begin{center}
\adjustbox{max width=0.95\linewidth}{
\begin{tabular}{cccccc}
\toprule
\multicolumn{2}{c}{\textbf{Attack}} & \textbf{BM} & \textbf{CCFD} & \textbf{MMNIST} & \textbf{CIFAR10} \\ \midrule

\multirow{2}{*}{\textbf{Query-Free}} & W/o \method & 0.1532 & 0.1325 & 0.6523 & 0.4875 \\
                                 & W/ \method  & 0.0221 & 0.0524 & 0.1514 & 0.1224 \\ \midrule
\multirow{2}{*}{\textbf{Genver}} & W/o \method & 0.1724 & 0.1566 & 0.7235 & 0.5243 \\
                                 & W/ \method  & 0.0132 & 0.0357 & 0.2432 & 0.1874 \\ \midrule
\multirow{2}{*}{\textbf{UIFV}}   & W/o \method & 0.1334 & 0.1224 & 0.6884 & 0.5432 \\
                                 & W/ \method  & 0.0187 & 0.0285 & 0.2412 & 0.2304 \\ \midrule
\multirow{2}{*}{\textbf{FIA}}    & W/o \method & 0.1893 & 0.1438 & 0.7543 & 0.5620 \\
                                 & W/ \method & 0.0321 & 0.0424 & 0.2565 & 0.2024 \\
\bottomrule
\end{tabular}
}
\label{tab:eval_privacy}
\end{center}
\end{table}

\begin{table}[!t]
\caption{The impact of privacy-aware training on VFL inference accuracy. Our mechanism only causes an average accuracy drop of 0.51\%.}
\begin{center}
\adjustbox{max width=0.95\linewidth}{
\begin{tabular}{cccccc}
\toprule
\multicolumn{2}{c}{\textbf{Metrics}} & \textbf{BM} & \textbf{CCFD} & \textbf{MMNIST} & \textbf{CIFAR10} \\ \midrule
\multirow{2}{*}{\textbf{Accuracy}}     & w/o \method & 74.97\% & 97.97\% & 97.26\% & 66.99\% \\
                                       & w/ \method & 74.24\% & 97.50\% & 96.67\% & 66.75\% \\
\bottomrule
\end{tabular}
}
\label{tab:accuracy}
\end{center}
\end{table}

\subsection{Real System Efficiency}
\label{subsec:system_efficiency}
We evaluate the latency overhead introduced by \method for auditing \data's untrusted inference.
Following Sec.~\ref{subsec:eo}, we denote the inference latencies of untrusted and trusted inference as $T_{un}$ and $T_{tr}$, respectively.
Tbl.~\ref{tab:system_efficiency} reports $T_{un}$ and $T_{tr}$ across five datasets.
At the optimal configuration $W_*$, the difference between $T_{un}$ and $T_{tr}$ is negligible.
It means \method ensures the inference trustworthiness of \data without introducing additional system latency.



\begin{table}[!t]
\caption{The running latency in untrusted ($T_{un}$) and trusted inference ($T_{tr}$).
Our mechanism only causes an average latency fluctuation of 0.56\% ($\frac{|T_{un}-T_{tr}|}{T_{un}}$).
}
\begin{center}
\adjustbox{max width=0.95\linewidth}{
\begin{tabular}{ccccccc}
\toprule
\textbf{Latency} & \textbf{BM} & \textbf{CCFD} & \textbf{MMNIST} & \textbf{CIFAR10} & \textbf{KDD-CUP} \\ \midrule
$T_{un}$ & 1.68 ($\pm$0.02) & 7.23 ($\pm$0.05) & 2.26 ($\pm$0.01) & 2.67 ($\pm$0.04) & 52.33 ($\pm$0.02) \\ \midrule
$T_{tr}$  & 1.68 ($\pm$0.02) & 7.27 ($\pm$0.05) & 2.25 ($\pm$0.01) & 2.68 ($\pm$0.03) & 51.87 ($\pm$0.05) \\ 
\bottomrule
\end{tabular}
}
\label{tab:system_efficiency}
\end{center}
\end{table}

\noindent \textbf{Ablation of Pipeline Acceleration.}
To validate the effectiveness of pipeline acceleration (in Sec.~\ref{subsec:efficiency_schedule}), we compared the validation latency of two variants of \method: with and without pipeline acceleration over different $W$. 
In each case, we find the optimal block size using the optimization strategy mentioned in Eq.~\ref{eq:optimal_block}.
Per Amdahl's Law~\cite{Amdahl67}, we focus on the optimized component's performance.
Tbl.~\ref{tab:cloud_improvement} presents the speedup ratio of \coord inference before and after applying pipeline acceleration.
Results show a consistent increase in speed ratio with larger $W$, with MMNIST and CIFAR-10 achieving significantly higher acceleration (averagely 40$\times$) compared to BM, CCFD, and KDD-CUP (averagely 6$\times$).
This is attributed to the larger parameter sizes of CNN and VGG16 used for MMNIST and CIFAR-10, compared to the FCNN used in the other datasets.
As a result, \coord-side inference accounts for a greater portion of total latency, amplifying the benefits of pipeline acceleration.
These observations suggest that \method is likely to achieve even larger efficiency gains when deployed on larger models.

\begin{table}[!t]
\footnotesize
\caption{The speedup ratio of \coord inference before and after pipeline acceleration.
The average value under $W_*$ is 4.38$\times$.
}
\begin{center}
\begin{tabular}{cccccc}
\toprule
\textbf{Dataset} & \textbf{$W=W_*$} & \textbf{$W=5\%$} & \textbf{$W=10\%$} & \textbf{$W=25\%$} & \textbf{$W=50\%$} \\ \midrule
BM      & 5.54$\times$ & 2.46$\times$ & 4.19$\times$ & 5.25$\times$ & 6.02$\times$ \\ \midrule
CCFD    & 8.51$\times$ & 3.46$\times$ & 8.32$\times$ & 9.42$\times$ & 13.9$\times$ \\ \midrule
MMNIST  & 1.95$\times$ & 38.2$\times$ & 68.3$\times$ & 43.9$\times$ & 109.5$\times$ \\ \midrule
CIFAR10 & 1.34$\times$ & 16.5$\times$ & 24.1$\times$ & 30.8$\times$ & 64.2$\times$ \\ \midrule
KDD-CUP & 4.54$\times$ & 1.88$\times$ & 3.62$\times$ & 5.74$\times$ & 7.85$\times$ \\ 
\bottomrule
\end{tabular}
\label{tab:cloud_improvement}
\end{center}
\end{table}

\subsection{Scalability to More Participants}
\label{subsec:scalability}

One potential concern of \method is the scalability to more participants.
While the typical VFL setting involves one \data and one \task (Sec.~\ref{subsec:vfl}), practical scenarios may involve multiple \data.
To assess overhead under such conditions, we vary the number of \data to $3$, $5$, $7$, and $9$, and measure the corresponding system latency.
For each \data, the malicious tampering ratio $K$ is randomly selected between $1\%$ and $100\%$, and a fixed $W_*$ is used for detection.
Tbl.~\ref{tab:scalability} reports the system latency over the single-party case (only one \data).
On average, the detection time across multiple \data increases by only 5\%, indicating that \method's overhead remains nearly constant as the number of \data grows.
These results indicate that \method scales effectively and is well-suited to multi-party VFL scenarios.


\begin{table}[!t]
\footnotesize
\caption{Average validation time of multiple \data. We report the ratio over the inference time of a single \data. 
}
\begin{center}
\adjustbox{max width=0.95\linewidth}{
\begin{tabular}{cccccc}
\toprule
\textbf{Dataset} & \textbf{Party=1} & \textbf{Party=3} & \textbf{Party=5} & \textbf{Party=7} & \textbf{Party=9} \\ \midrule
BM      & $1.00\times$ & $1.03\times$ & $1.30\times$ & $1.18\times$ & $1.15\times$ \\ \midrule
CCFD    & $1.00\times$ & $1.03\times$ & $1.06\times$ & $1.01\times$ & $1.03\times$ \\ \midrule
MMNIST  & $1.00\times$ & $1.09\times$ & $1.13\times$ & $0.97\times$ & $0.96\times$ \\ \midrule
CIFAR10 & $1.00\times$ & $0.98\times$ & $0.92\times$ & $0.92\times$ & $0.96\times$ \\ \midrule
KDD-CUP & $1.00\times$ & $1.02\times$ & $1.05\times$ & $1.09\times$ & $1.04\times$ \\
\bottomrule
\end{tabular}
}
\label{tab:scalability}
\end{center}
\end{table}

\section{Discussion}
\label{sec:dis}

\myparagraph{Defense Scenarios.}
\method targets the common online setting where inference throughput is extremely high and ``always-on'' per-inference verification is not a realistic default.
Our defense is designed to be recall-first: \method raises a high-confidence alarm without impacting the inference execution, and this alarm can trigger an escalation workflow to produce fine-grained evidence.
Hence, fine-grained detection is compatible with \method. It is a complementary capability that can be layered on top when the operating context demands it.


\myparagraph{Real-World Deployments.}
Our current implementation of \method is based on Intel SGX~\cite{CostanD16}.
However, \method can be easily extended to other TEE hardware platforms, such as ARM TrustZone~\cite{Winter08} and AMD Secure Encrypted Virtualization~\cite{amdsev}.
It is because \method relies on the confidentiality and integrity guarantees provided by the TEE runtime rather than on specific hardware infrastructure.
As a result, \method can be rapidly deployed and effectively perform auditing, even in VFL inference tasks involving heterogeneous systems among VFL participants.
In our experiments, we use the Intel i7-9700KF CPU, which supports an EPC size of up to 128MB (without considering page swapping) for SGX.
Existing server-level Intel Xeon processors provide an EPC size of up to 512GB and additional cores~\cite{intel-sgx-support}, enabling inference auditing for models with larger parameter sizes.
In VFL settings, the participants are usually enterprises or institutions with richer computing resources, making TEE-based deployment a practical and feasible solution.

\myparagraph{Impact on Numerical Precision.}
The transition between TEE-COO computing environments can introduce variations in numerical precision.
For instance, the same data may be subject to different precision truncation in TEE and GPU environments.
\method's data/model validation and inference consistency verification rely on hash computations, which may be sensitive to minor numerical discrepancies.
As a result, small changes in precision can lead to differing hash values for the same inputs.
To ensure \method's inference auditing be reliable, the validation results are lossless--i.e., any detected inconsistency must originate from malicious tampering by \data, rather than benign variations in numerical precision.
The results in Tbl.~\ref{tab:vefia_defense} show that \method achieves a PPV of $100\%$, indicating all inferences flagged as malicious are true positives, with no false positives.
This result validates that \method provides effective and reliable detection without compromising numerical precision.

\myparagraph{Collusion.}
\method follows the standard VFL assumption that \coord is semi-honest and does not collude with any participant~\cite{YangLCT19,HeZL19,GaoZ23}.
This is consistent with prior VFL literature and practical deployments, where collusion would severely damage the coordinator's business credibility and contractual compliance.
Under full collusion between \coord and \data, the adversary may try to influence the trusted validation path and thus undermine inconsistency-based auditing; therefore, such full collusion is outside the scope of our current threat model.
That said, \method can be strengthened by reducing the single-point trust on a single coordinator.
A promising future direction is to adopt a multi-auditor architecture: the trusted validation path is executed by multiple independent coordinators (or auditing services), and \task cross-checks their attested outputs.
As long as at least one auditor remains non-colluding, the adversary cannot simultaneously forge all trusted validation results, enabling robust auditing under partial collusion.
We leave the full system design and evaluation of this multi-auditor extension to future work.

\section{Related Work}

\myparagraph{TEE for FL.}
PPFL utilizes TEE on both client training and server aggregation against attacks on gradients and models~\cite{MoHKMPK21}.
Olive uses TEE to implement model aggregation on untrusted servers and designs a new aggregation algorithm to prevent memory access pattern leakage~\cite{Kato0Y23}.
Some other studies~\cite{ZhangWCHM21,ZhaoJFWSL22} have also shown that TEE-enabled FL has great potential.
However, these works \textit{only consider HFL}.
This paper focuses on how to leverage TEE in VFL to audit the trustworthiness of the \data's inference process.

\myparagraph{TEE-based secure inference.}
TEE-based secure inference generally falls into two categories.
The first, TEE-shielding approaches~\cite{Hanzlik0G0A0F21,LeeLPLLLXXZS19}, execute the entire inference process within a secure enclave.
The second type~\cite{ShenQJWWCZWCLZC22,abs-1912-03485,Zhangteeslice2024} is the partitioning-based method, which executes a portion of model layers within the secure enclave.
However, these methods can only guarantee the inference integrity but cannot ensure the authenticity of the data and model.


\myparagraph{FL Auditing.}
Chang et al. propose novel and efficient membership inference attacks for auditing privacy risks in HFL~\cite{ChangEPS24}.
Zhang et al. propose a novel FL framework that incorporates a data integrity auditing function to evaluate the reputation of each participant~\cite{ZhangL24}.
These FL audit frameworks primarily \textit{focus on HFL} and aim to protect participant privacy during the \textit{training phase}~\cite{ChangEPS24,ZhangL24}.
In contrast, our work targets the joint inference phase of VFL, ensuring that the inference behavior of \data is executed as expected by \task through inference consistency audit.

\myparagraph{Zero-Knowledge Proofs for FL.}
Zero-knowledge proofs (ZKPs) have been explored in HFL mainly as a training-time verifiability layer~\cite{0001GLLM0Y23,ZhuWLOX24}.
Systems such as ACORN~\cite{0001GLLM0Y23} and RiseFL~\cite{ZhuWLOX24} integrate ZKPs into secure aggregation, thereby improving practicality.
Yet, their proving cost still scales with the model dimensionality and is typically non-trivial even for moderate-to-large updates~\cite{0001GLLM0Y23,ZhuWLOX24}.
These ZKP-based FL mechanisms would introduce prohibitive overhead and violate our zero-latency design goal.
Therefore, these mechanisms are therefore not suitable as baselines for VFL online inference auditing.

\section{Conclusion}
VFL is a key technology for distributed privacy-preserving AI;
however, existing research has failed to address the issue of whether inference is trustworthy, which poses a significant obstacle to the practical application of VFL.
\attack allows \data to disrupt the \task's joint inference with minimal tampering, making it difficult to detect.
\method enables \task to use the TEE-COO trusted collaborative inference to audit \data's inference as expected without extra system latency.
Finally, the experiments have demonstrated robust performance in terms of privacy protection and scalability.

\bibliographystyle{ACM-Reference-Format}
\bibliography{reference}

\appendix

\section{Background}
\label{app:background}
\subsection{HFL vs VFL}

\begin{figure}[htbp]
\centerline{\includegraphics[width=0.7\linewidth]{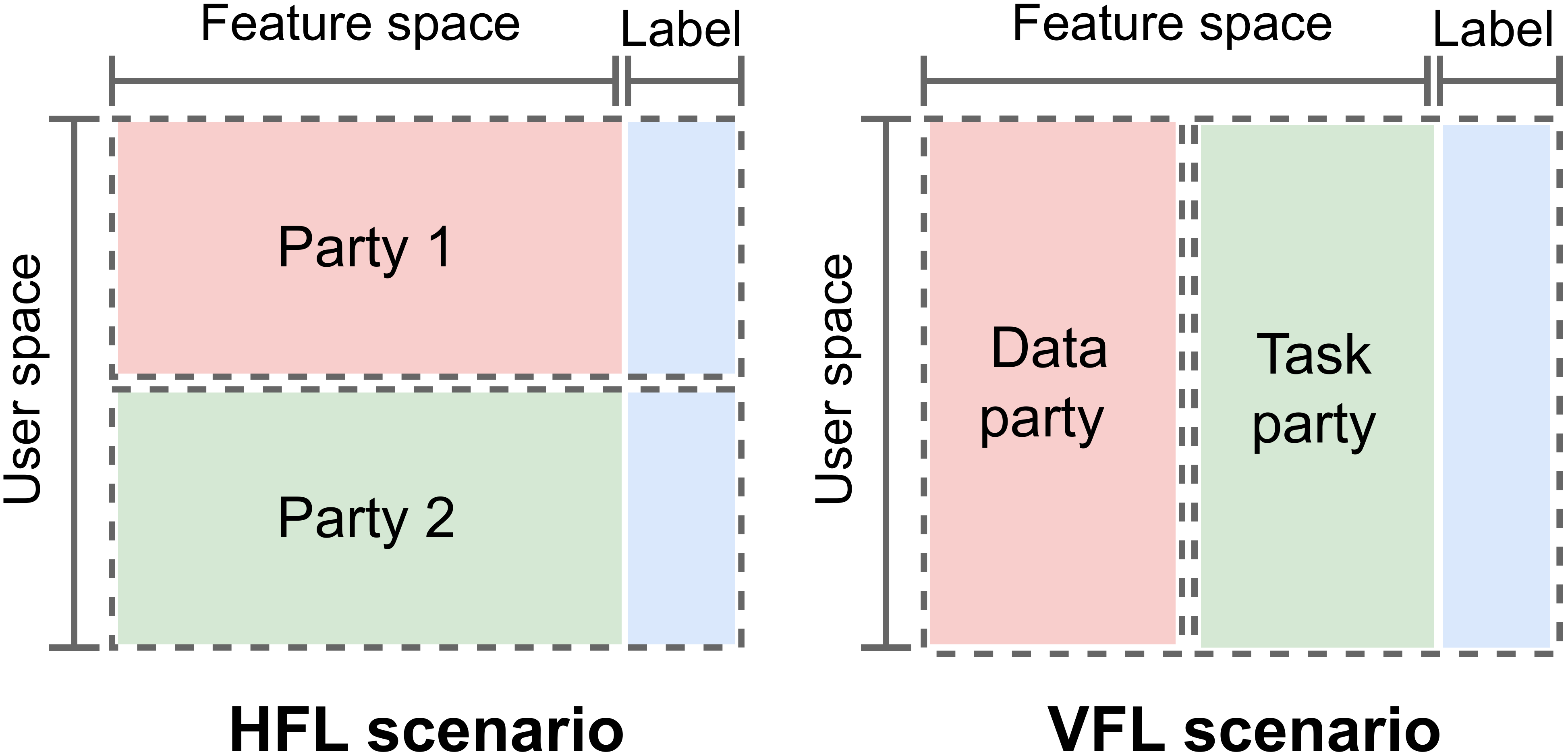}}
\caption{The scenario comparison between HFL and VFL.}
\label{fig:scenario}
\Description{}
\end{figure}

FL is categorized into Horizontal FL (HFL) and Vertical FL (VFL) \cite{YangLCT19}.
In HFL, each participant holds complete feature and label data for different users.
The aim of HFL is to collaboratively train a shared global model without sharing user data.
HFL only involves the training phase, but after training, each participant can personalize and fine-tune the local model for their own inference tasks.
In the inference, the parties of HFL do not rely on other parties to make predictions because each party has the whole feature.
In VFL, participants hold different feature sets for the same users, but only the task party has the labels.
Additionally, each participant possesses a portion of the model based on Split Neural Network (SplitNN)~\cite{abs-1812-00564}, making it impossible to complete training or inference independently.
Therefore, unlike HFL, VFL includes both a joint training phase (\train) and a joint inference phase (\infer).
Fig.~\ref{fig:scenario} shows the scenario differences between HFL and VFL.

\section{\attack's Design \& Evaluation}


\subsection{Attack Metrics}
\label{app:eval_metrics}

We use the following five metrics to evaluate \attack:

\begin{itemize}[leftmargin=*]
    \item \textit{Accuracy} measures the accuracy of \task's joint inference task.
    Recall that the adversary's goal is to reduce the accuracy by tampering with the inference results (Sec.~\ref{sec:threat_model}).
    A lower accuracy indicates a higher attack performance.
    \item \textit{Attack Success Rate (ASR)} measures the success rate of \attack to cause misclassification by \data.
    It is defined as the ratio of samples correctly classified before the attack but misclassified afterward.
    A higher ASR indicates a better attack performance.
    \item \textit{Positive Predictive Value (PPV)} measures the proportion of truly malicious inferences among those identified as malicious by the defense mechanism.
    True Positives (TP) represents the number of correctly identified malicious inferences.
    False Positive (FP) is the number of benign inferences incorrectly identified as malicious.
    PPV is computed as $PPV=\frac{TP}{TP+FP}$.
    \item \textit{True Positive Rate (TPR)} measures the proportion of correctly identified malicious inferences to all malicious inferences.
    Let False Negatives (FN) represent the number of malicious inferences that are incorrectly identified as benign.
    TPR is computed as $TPR=\frac{TP}{TP+FN}$.
    \item \textit{Negative Predictive Value (NPV)} measures the proportion of correctly identified benign inferences to all benign inferences.
    True Negatives (TN) is the total number of correctly identified benign inferences.
    NPV is computed as $NPV=\frac{TN}{TN+FN}$.
\end{itemize}

\subsection{Evaluation Setup}
\label{app:eval_setup}
\myparagraph{Dataset, Models, and Partitions.}
BM and CCFD are tabular datasets.
BM has 41,188 samples, 63 features, and 2 classes.
CCFD has 284,807 samples, 29 features, and two classes. 
MMNIST and CIFAR10 are image datasets.
MMNIST has 58,954 medical images (64$\times$64) of 6 classes, and CIFAR10 has 50,000 images (3$\times$32$\times$32) of 10 classes.
KDD-CUP is a large-scale tabular dataset with 4,898,431 samples, 33 features, and 5 classes.
Tbl.~\ref{tab:dataset} shows the feature split rule. 
Due to label imbalance in BM and CCFD, we applied the SMOTE oversampling technique to enhance the sample set. 
We use a 7:3 training-to-inference ratio.
The model architectures for VFL training and inference are shown in Tbl.~\ref{tab:model_split}.
For tabular datasets (BM and CCFD), we use a six-layer fully connected neural network (FCNN), where the bottom model consists of four FCNN layers and the top model comprises three FCNN layers.
For MMNIST, the bottom model is a three-layer convolutional neural network (CNN), while the top model is a one-layer FCNN.
For CIFAR-10, we use VGG16 as the bottom model and a three-layer FCNN as the top model.
Tbl.~\ref{tab:model_split1} presents the model partitioning scheme used in TEE-COO collaborative inference.

\begin{table}[htbp]
\caption{Data partition in our evaluation.}
\begin{center}
\begin{tabular}{cccc}
\toprule
\textbf{Dataset} & \textbf{Type} & \textbf{Sample} & \textbf{Feature} \\ \midrule
BM      & Tabular     & 82,376    & \task: 32, \data: 31 \\ \midrule
CCFD    & Tabular     & 569,614   & \task: 15, \data: 14 \\ \midrule
MMNIST  & Image       & 58,954    & \task,\data: 32$\times$64  \\ \midrule
CIFAR10 & Image       & 50,000    & \task,\data: 3$\times$16$\times$32 \\ \midrule
KDD-CUP & Tabular     & 4,898,431 & \task: 15, \data: 15 \\ \bottomrule
\end{tabular}
\label{tab:dataset}
\end{center}
\end{table}

\begin{table}[htbp]
\caption{Model partition for VFL training and inference.}
\begin{center}
\begin{tabular}{cccc}
\toprule
\textbf{Dataset} & \textbf{Bottom Model} & \textbf{Top Model} \\ \midrule
BM               & FCNN-3                & FCNN-4             \\ \midrule
CCFD             & FCNN-3                & FCNN-4             \\ \midrule
MMNIST           & CNN-3                 & FCNN-1             \\ \midrule
CIFAR10          & VGG16-13              & FCNN-3             \\ \midrule
KDD-CUP          & FCNN-3                & FCNN-4             \\
\bottomrule
\end{tabular}
\label{tab:model_split}
\end{center}
\end{table}

\begin{table}[htbp]
\caption{Model partition for TEE-COO collaborative inference.}
\begin{center}
\begin{tabular}{cccc}
\toprule
\textbf{Dataset} & \textbf{Shallow Model} & \textbf{Deep Model} \\ \midrule
BM               & FCNN-1                & FCNN-2            \\ \midrule
CCFD             & FCNN-1                & FCNN-2            \\ \midrule
MMNIST           & CNN-1                 & CNN-2             \\ \midrule
CIFAR10          & VGG16-1               & VGG16-12          \\ \midrule
KDD-CUP          & FCNN-1                & FCNN-2            \\
\bottomrule
\end{tabular}
\label{tab:model_split1}
\end{center}
\end{table}


\subsection{Attack Evaluations}
\label{app:attack_eval}
\begin{figure}[htbp]
    \centering
    \begin{subfigure}{0.48\linewidth}
        \centering
        \includegraphics[width=1.0\linewidth]{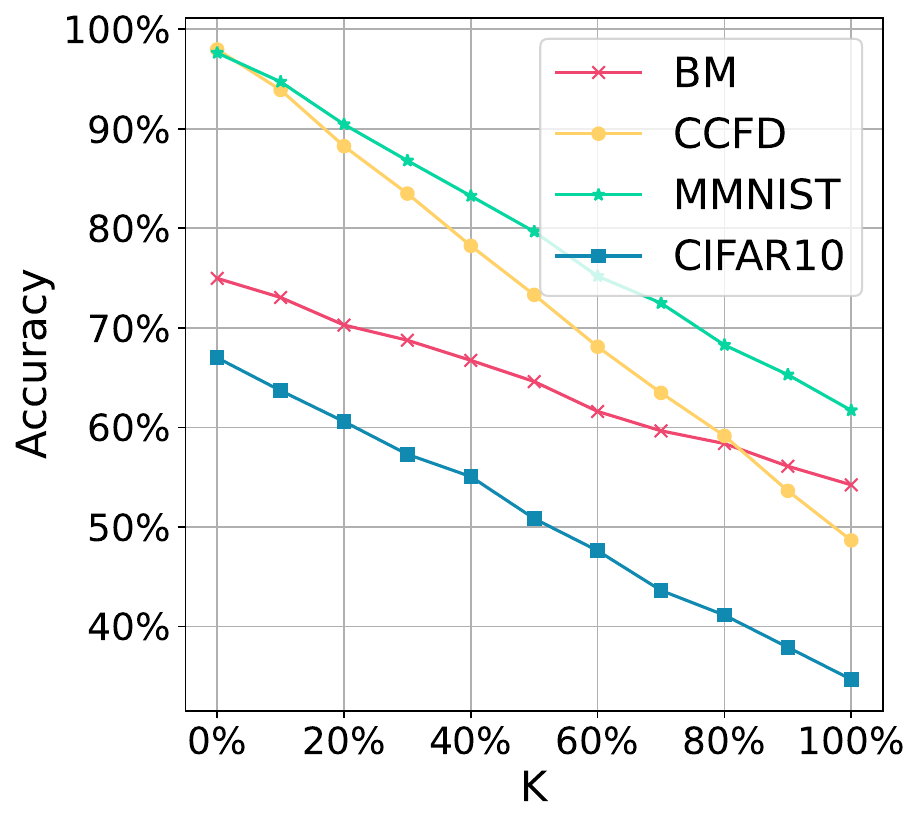}
        \caption{Accuracy vs. $K$}
        \label{fig:attack_k}
    \end{subfigure}
    \begin{subfigure}{0.48\linewidth}
        \centering
        \includegraphics[width=1.0\linewidth]{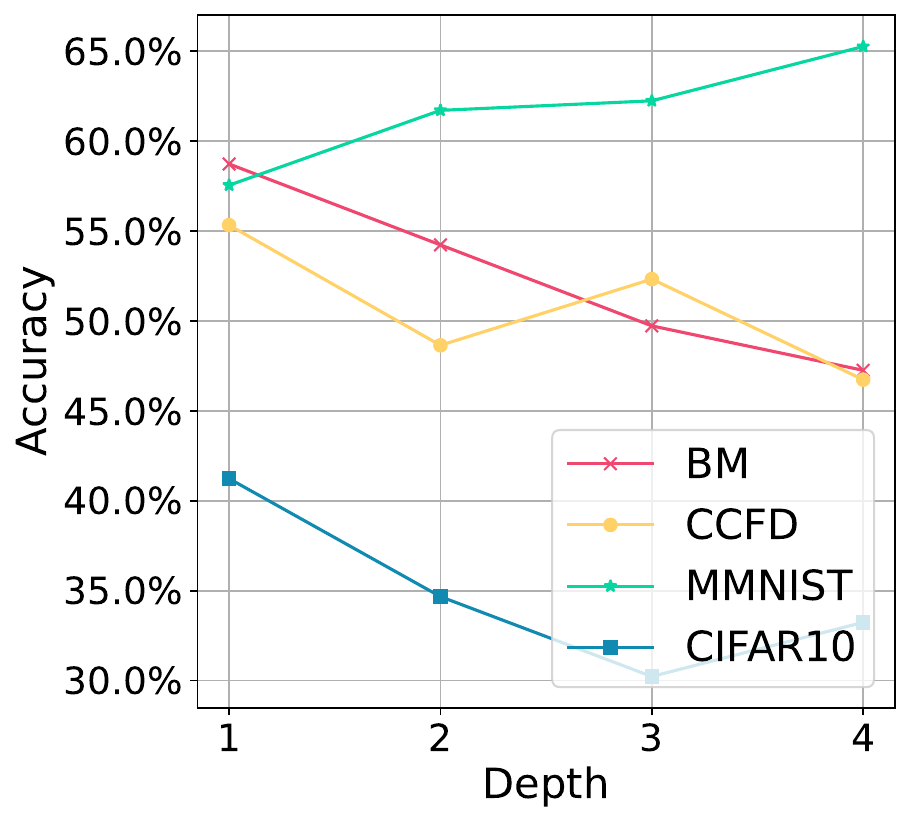}
        \caption{Accuracy vs. $g_s$ depth}
        \label{fig:attack_gs_acc}
    \end{subfigure}\hfill
    \caption{The changes of accuracy as $K$ or $g_s$ depth increases.}
    \Description{}
\end{figure}

\begin{figure}[htbp]
\centerline{\includegraphics[width=0.65\linewidth]{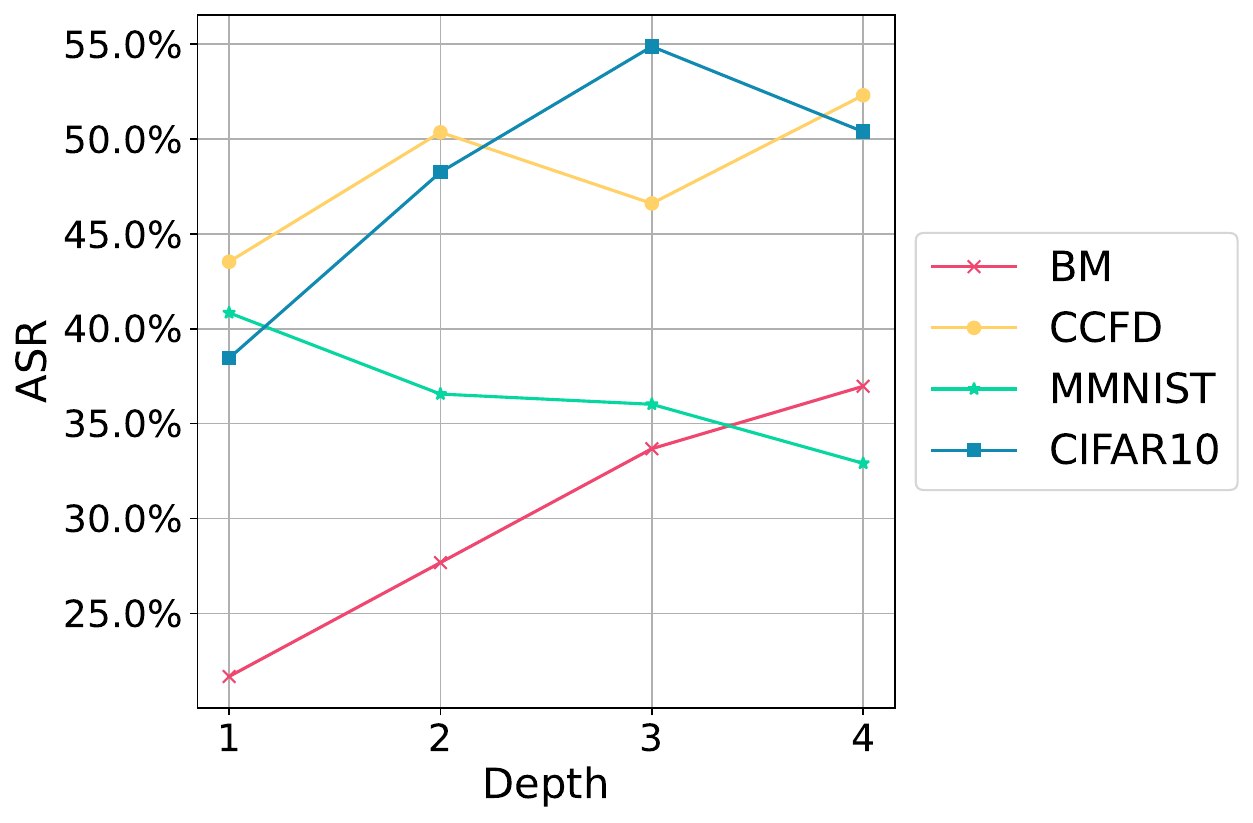}}
\caption{The changes of ASR as $g_s$ depth increases.}
\label{fig:attack_gs_asr}
\Description{}
\end{figure}

We progressively increased the $K$ during the deployment of the \attack to evaluate the attack effectiveness, as shown in Fig.~\ref{fig:attack_k}.
As observed, accuracy gradually decreases with the increase in $K$.
The reduction in accuracy remains consistent because the ASR of \attack is independent of the $K$.
We further examined the effect of different depths of $g_s$ on accuracy and ASR when $K=100\%$.
As shown in Fig.~\ref{fig:attack_gs_acc} and Fig.~\ref{fig:attack_gs_asr}, across four datasets, when the depth of $g_s$ matches the depth of $g$ in Tbl.~\ref{tab:model_split}, \attack achieves the greatest reduction in accuracy and greatest ASR.


\section{\method's Design \& Evaluations}

\subsection{TEE-COO partition}
\label{app:tee_coo}
\begin{figure}[htbp]
\centerline{\includegraphics[width=0.75\linewidth]{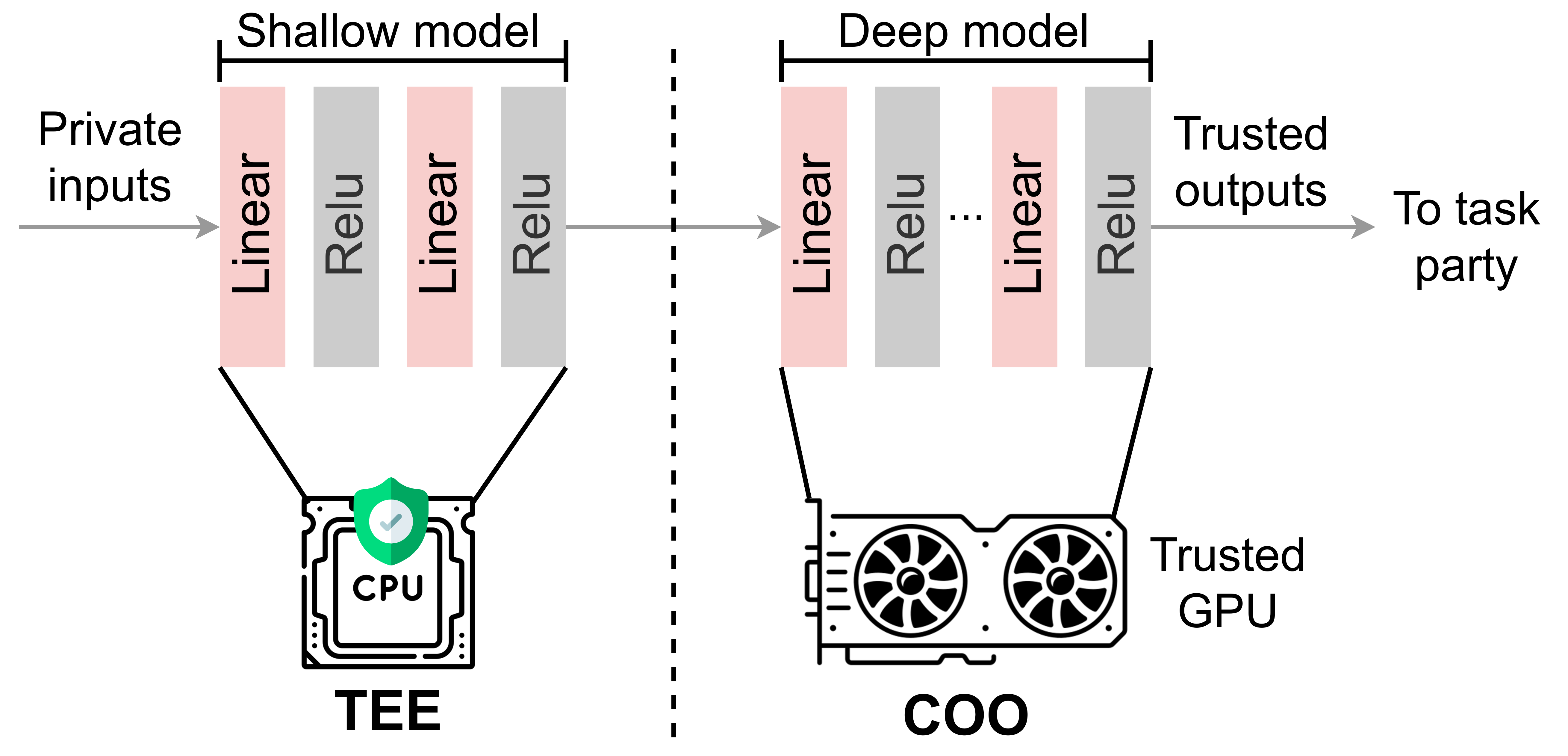}}
\caption{TEE-COO collaborative inference using a FCNN as an example.}
\label{fig:model_partition}
\Description{}
\end{figure}

Inspired by the prior cloud-edge collaboration mechanism~\cite{mo2020darknetz,abs-1912-03485,LiuGZZJLLZ24}, we designed a model partition mechanism for collaborative inference between the \data's local TEE and the \coord's trusted GPU, as shown in Fig.~\ref{fig:model_partition}.

\subsection{Confidential Random Sampling Validation}
\label{app:crsv}
\begin{figure}[!htbp]
\centerline{\includegraphics[width=0.9\linewidth]{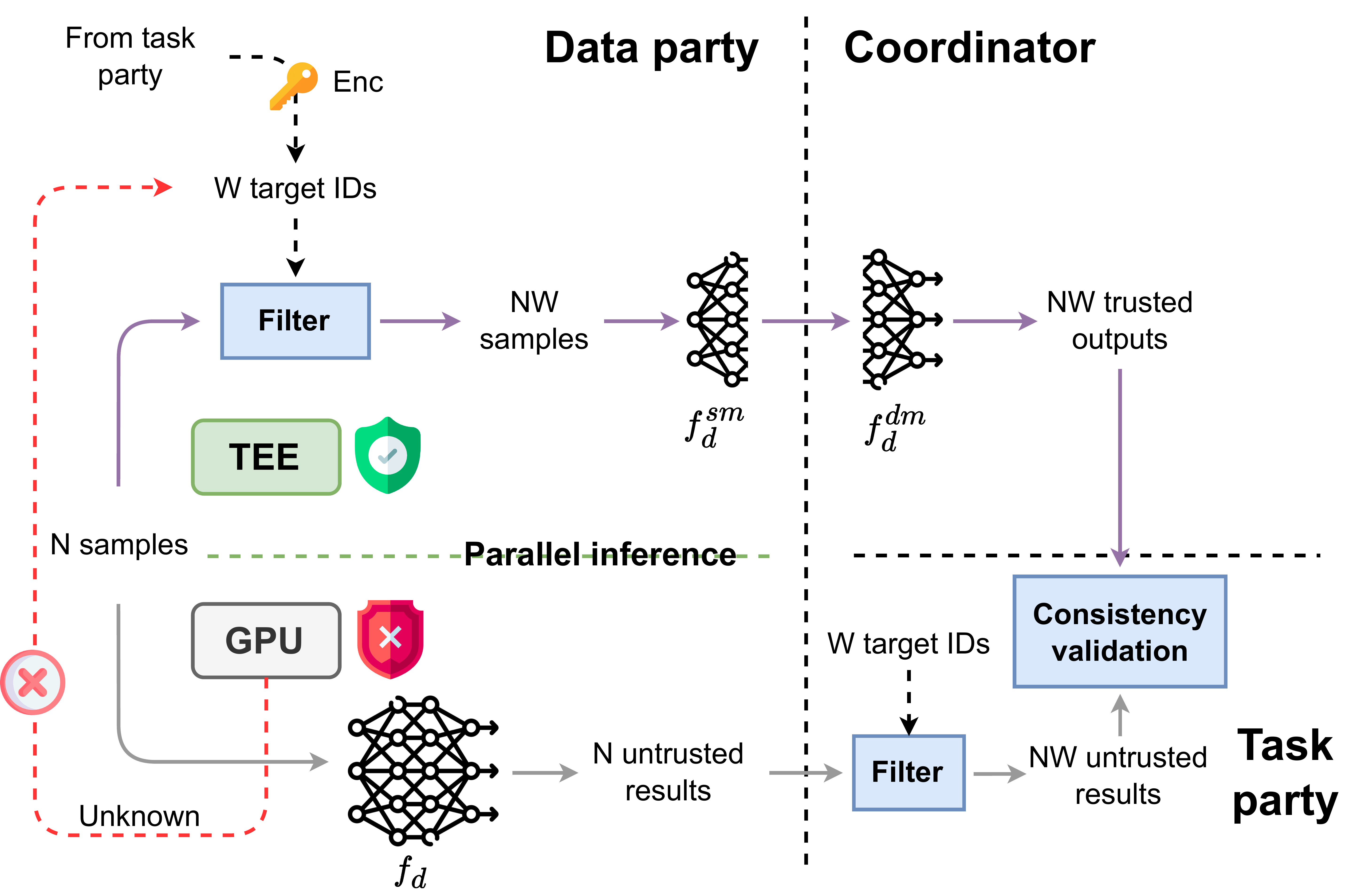}}
\caption{Confidential random sampling validation during online inference.}
\label{fig:crsv}
\Description{}
\end{figure}

We design a malicious inference detection mechanism \method based on confidential random sampling validation as shown in Fig.~\ref{fig:crsv} to achieve fast validation on large-scale inference queries.

\subsection{Proof of Thm.~\ref{theorem:bound}}
\label{app:proof}
We aim to derive the following lower bound for the reconstruction error:
\begin{equation}
    \mathbb{E}(||x_d-\mathcal{A}(\hat z_d)||_q/m)\ge \frac{e^\frac{2}{m}h(x_d)}{2\pi e}e^{-\frac{2}{m}I(x_d;\hat z_d)}
\end{equation}
Here: $x_d\in \mathbb{R}^m$ is the samples of \data; $\hat z_d$ represents the available intermediate information of \coord; $\mathcal{A}$ is any MoIA attack launched by \coord based on $\hat z_d$; $I(x_d;\hat z_d)$ is the mutual information between $x_d$ and $\hat z_d$; $h(x_d)$ is the differential entropy of $x_d$.
Before the derivation, we introduce the key theoretical tools.

\myparagraph{Differential Entropy.}
In information theory, differential entropy is the continuous counterpart of Shannon entropy, 
quantifying the uncertainty or randomness of a continuous random variable.
For a continuous random vector $X\in \mathbb{R}^m$ with probability density function $p(x)$, the differential entropy is defined as:
\begin{equation}
    h(X)=-\int_{\mathbb{R}^m}p(x)\log p(x)dx
\end{equation}
Differential entropy measures how ``spread out'' a distribution is in space. Unlike discrete entropy, differential entropy can take negative values and is not invariant under changes in the coordinate system. However, it still follows many key properties of discrete entropy, making it useful in signal processing, communication theory, and privacy analysis:
\begin{itemize}
    \item \textit{Translation Invariance.} If $X$ is shifted by a constant vector $a$, the entropy remains unchanged: $h(X+a)=h(X)$.
    \item \textit{Maximum Entropy Property.} Among all continuous distributions with a fixed covariance matrix, the multivariate Gaussian distribution has the highest differential entropy: $h(X)\le \frac{m}{2}\log(\frac{2\pi e}{m}\mathbb{E}[||X||^2])$.
    \item \textit{Relation to Mutual Information.} The mutual information between two variables is given by: $I(X;Y)=h(X)-h(X|Y)$.
\end{itemize}

\myparagraph{Derivation of the Lower Bound.}
We now proceed with the derivation. We define the MoIA reconstruction error as:
\begin{equation}
    \mathbb{E}[||x_d-\mathcal{A}(\hat z_d)||_p]=\mathbb{E}[||\epsilon||_p]
\end{equation}
Applying the maximum entropy property, we obtain:
\begin{equation}
\label{eq:mep}
    \mathbb{E}[||\epsilon||_p/m]\ge \frac{1}{2\pi e}e^{\frac{2}{m}h(\epsilon)}
\end{equation}
From an information-theoretic perspective, the conditional entropy $h(\epsilon|\hat z_d)$ represents the remaining uncertainty in $\epsilon$ after knowing $\hat z_d$, whereas $h(\epsilon)$ represents the total uncertainty in $\epsilon$. Since knowing $\hat z_d$ provides additional information about $\epsilon$, the uncertainty cannot increase, i.e., $h(\epsilon|\hat z_d) \le h(\epsilon)$. This indicates that conditioning on $\hat z_d$ reduces or at least maintains the same level of uncertainty in $\epsilon$. Therefore, Eq.~\ref{eq:mep} can be rewritten as:
\begin{equation}
\label{eq:mep1}
    \mathbb{E}[||\epsilon||_p]\ge \frac{m}{2\pi e}e^{\frac{2}{m}h(\epsilon|\hat z_d)}
\end{equation}
Since $\mathcal{A}(\hat z_d)$ is uniquely determined by $\hat z_d$, then from translation invariance we have:
\begin{equation}
    h(\epsilon|\hat z_d)=h(x_d-\mathcal{A}(\hat z_d)|\hat z_d)=h(x_d|\hat z_d)
\end{equation}
By using the relationship between differential entropy and mutual information, we obtain:
\begin{equation}
    h(x_d|\hat z_d)=h(x_d)-I(x_d;\hat z_d)
\end{equation}
Eq.~\ref{eq:mep1} can be rewritten as:
\begin{align}
    \mathbb{E}(||x_d-\mathcal{A}(\hat z_d)||_q/m)&\ge \frac{1}{2\pi e}e^{\frac{2}{m}(h(x_d)-I(x_d;\hat z_d))}\\
    &=\frac{e^\frac{2}{m}h(x_d)}{2\pi e}e^{-\frac{2}{m}I(x_d;\hat z_d)}
\end{align}
where the last equality is a consequence of Eq.~\ref{eq:lower_bound}.

\subsection{Defense Metrics}
\label{app:defense_metrics}

We used the two metrics to evaluate the privacy protection of \method.

\begin{itemize}
    \item \textit{Hitting Rate (HR)}:
    Following prior work~\cite{LiuF00024,abs-2404-15821}, we use HR to quantify the similarity between original and reconstructed tabular data (BM and CCFD).
    HR measures the proportion of original data that is successfully matched by the reconstructed data. Lower HR values indicate stronger resistance to inversion attacks.
    We set a safety threshold as 0.09, below which inversion is deemed unsuccessful~\cite{abs-2404-15821}.
    \item \textit{Structural Similarity Index Measure (SSIM)}~\cite{WangBSS04}:
    For image data (MMNIST and CIFAR10), we use SSIM to evaluate perceptual similarity between original and reconstructed images.
    SSIM ranges from 0 to 1, with lower values indicating higher distortion and thus stronger defense against inversion.
    Following prior work~\cite{HeZL19}, we set a safety threshold as 0.3, below which reconstructed images are deemed unrecognizable.
\end{itemize}

\subsection{Model Inversion Attacks}
\label{app:moia}

In Sec.~\ref{subsec:privacy_eval}, we employ the SOTA inference-phase model inversion attack to evaluate the privacy protection of \method.
Notably, during \train, \coord functions solely as an aggregator, combining model outputs from multiple parties.
It is only during \infer that \data outsources the partial bottom model to \coord for inference.
Consequently, \coord can only conduct attacks during \infer.
\begin{itemize}
    \item Query-free model inversion attack.
    He et al.~\cite{HeZL19} first introduced model inversion attacks, which enable the reconstruction of \data's original data in query-free collaborative inference.
    Query-free setting leverages a shadow dataset to train a surrogate model that approximates the functionality of $f_d^{sm}$.
    Regularized maximum likelihood estimation is then applied to the surrogate model to recover sensitive inputs.
    \item Ginver~\cite{YinZZLYCH23}.
    Compared to query-free model inversion attack, Ginver needs a stronger attacker.
    In this setting, \coord can query $f_d^{sm}$ using a shadow dataset and then train an inversion model $f^{-1}$ to reconstruct the original data.
    \item UIFV~\cite{abs-2406-12588}.
    UIFV is another query-based attack that utilizes a data generator to synthesize a set of fake samples, which are then used to query $f_d^{sm}$ and extract intermediate features.
    Subsequently, UIFV trains an inversion model to reconstruct the original data.
    \item FIA~\cite{LuoWXO21}.
    FIA assumes that the attacker has access to the parameters of $f_d^{sm}$ and uses the generative regression network along with the maximum a posteriori estimation method to infer the inputs of \data.
\end{itemize}

\subsection{Detection Performance Validation}
\label{app:detection_performance}
We validated PPV, TPR, and NPV of \method within the optimal $W_*$ in Sec.~\ref{subsec:main_eval}.
To further validate the detection performance of \method, we test it in the different $W$. Tbl.~\ref{tab:vfedti_defense_w} shows that under different validation ratios, \method can obtain 100\% PPV, TPR, and NPV.

\begin{table}[!t]
\caption{The defense performance of \method measured by PPV, TPR, and NPV.}
\begin{center}
\begin{tabular}{cccccc}
\toprule
\textbf{$W$} & \textbf{Metrics} & \textbf{BM} & \textbf{CCFD} & \textbf{MMNIST} & \textbf{CIFAR10} \\ \midrule

\multirow{3}{*}{\textbf{$W=W_*$}}  & PPV & 100.0\% & 100.0\% & 100.0\% & 100.0\% \\
                                   & TPR & 100.0\% & 100.0\% & 100.0\% & 100.0\% \\
                                   & NPV & 100.0\% & 100.0\% & 100.0\% & 100.0\% \\ \midrule

\multirow{3}{*}{\textbf{$W=10\%$}} & PPV & 100.0\% & 100.0\% & 100.0\% & 100.0\% \\
                                   & TPR & 100.0\% & 100.0\% & 100.0\% & 100.0\% \\
                                   & NPV & 100.0\% & 100.0\% & 100.0\% & 100.0\% \\ \midrule
\multirow{3}{*}{\textbf{$W=30\%$}} & PPV & 100.0\% & 100.0\% & 100.0\% & 100.0\% \\
                                   & TPR & 100.0\% & 100.0\% & 100.0\% & 100.0\% \\
                                   & NPV & 100.0\% & 100.0\% & 100.0\% & 100.0\% \\ \midrule
\multirow{3}{*}{\textbf{$W=50\%$}} & PPV & 100.0\% & 100.0\% & 100.0\% & 100.0\% \\
                                   & TPR & 100.0\% & 100.0\% & 100.0\% & 100.0\% \\
                                   & NPV & 100.0\% & 100.0\% & 100.0\% & 100.0\% \\
\bottomrule
\end{tabular}
\label{tab:vfedti_defense_w}
\end{center}
\end{table}

\section*{Open Science}

We follow the open science principles encouraged by the community.
To foster transparency, we release the implementation of \method in this anonymous link \url{https://anonymous.4open.science/r/VeFIA}





\end{document}